\documentclass[a4paper,fleqn]{cas-dc}

\usepackage{paralist}
\usepackage{graphicx}
\usepackage[utf8]{inputenc}
\usepackage[sort&compress,numbers]{natbib}
\usepackage{enumitem}
\usepackage{multirow}
\usepackage{threeparttable}
\usepackage{caption}
\usepackage[toc,page]{appendix}
\usepackage{amssymb,amsmath} 
\usepackage{longtable}
\usepackage{arydshln}
\newcolumntype{L}[1]{>{\raggedright\let\newline\\\arraybackslash\hspace{0pt}}p{#1}}
\usepackage{framed}
\usepackage{listings}
\usepackage{placeins}
\usepackage{changepage}
\usepackage{algpseudocode}
\usepackage{algorithm}
\floatname{algorithm}{Pseudocode}
\algnewcommand\algorithmicLet{\textbf{Let}}
\algnewcommand\Let{\item[\algorithmicLet]}
\algnewcommand\algorithmicSet{\textbf{Set}}
\algnewcommand\Set{\item[\algorithmicSet]}
\algnewcommand\algorithmicForeach{\textbf{foreach}}
\algdef{S}[FOR]{Foreach}[1]{\algorithmicForeach\ #1\ \algorithmicdo}

\begin{document}
\let\WriteBookmarks\relax
\def\floatpagepagefraction{1}
\def\textpagefraction{.001}
\shortauthors{Tran et~al.}

\shorttitle{A proposal and assessment of an improved heuristic for the Eager Test smell detection}

\title[mode=title]{A proposal and assessment of an improved heuristic for the Eager Test smell detection}

\author[1]{Huynh Khanh Vi Tran}[type=editor]
\cormark[1]
\ead{huynh.khanh.vi.tran@bth.se}

\address[1]{Blekinge Institute of Technology, Department of Software Engineering, SE-37179, Karlskrona, Sweden}

\author[1]{Nauman bin Ali}
\ead{nauman.ali@bth.se}

\author[1]{Michael Unterkalmsteiner}
\ead{michael.unterkalmsteiner@bth.se}

\author[1]{J\"urgen B\"orstler}
\ead{jurgen.borstler@bth.se}

\cortext[cor1]{Corresponding author}

\begin{abstract}
\textbf{Context}: The evidence for the prevalence of test smells at the unit testing level has relied on the accuracy of detection tools, which have seen intense research in the last two decades. The Eager Test smell, one of the most prevalent, is often identified using simplified detection rules that practitioners find inadequate.

\textbf{Objective}: We aim to improve the rules for detecting the Eager Test smell.

\textbf{Method}: We reviewed the literature on test smells to analyze the definitions and detection rules of the Eager Test smell. We proposed a novel, unambiguous definition of the test smell and a heuristic to address the limitations of the existing rules. We evaluated our heuristic against existing detection rules by manually applying it to 300 unit test cases in Java.

\textbf{Results}: Our review identified 56 relevant studies. We found that inadequate interpretations of original definitions of the Eager Test smell led to imprecise detection rules, resulting in a high level of disagreement in detection outcomes. Also, our heuristic detected patterns of eager and non-eager tests that existing rules missed.

\textbf{Conclusion}: Our heuristic captures the essence of the Eager Test smell more precisely; hence, it may address practitioners’ concerns regarding the adequacy of existing detection rules.

\end{abstract}


\maketitle


\section{Introduction} \label{sec:introduction}
Automated testing has received more attention in recent years as it enables continuous software engineering~\cite{JabbariAPT18}. Test code serves as the backbone of automated software testing. Therefore, sufficient test code quality is required to maintain confidence in automated testing. However, developing and maintaining high-quality test code has proven to be non-trivial~\cite{garousi2018smells}.

Several guidelines and practices have been proposed to achieve high-quality test code at the unit testing level.
Like the smells of the production code, improper testing practices lead to \emph{test smells}, ``indicating trouble in test code'', as defined by van Deursen et al.~\cite{van2001refactoring}.
A common test smell that has been intensively studied in the literature is the Eager Test smell, which is commonly referred as a test method checking several methods of the object under test.
Many detection tools (TestQ~\cite{breugelmans2008testq}, Test Smell Detector~\cite{Bavota2015Are}, TASTE~\cite{Palomba2018Automatic}, SOCRATES~\cite{Bleser2019SoCRATES}, DARTS~\cite{Lambiase2020Just}, tsDetect~\cite{Peruma2020TsDetect}, and JNose Test~\cite{Virginio2019On}) were developed to detect test smells, including the Eager Test smell.
Several studies~\cite{Bavota2015Are, Tufano2016An, Van2006Characterizing, Martins2023On, Damasceno2022Analyzing, Bavota2012empirical, Spadini2018On} used these tools to analyze the prevalence of the Eager Test smell in practice and to investigate the impact of the test smell on multiple quality-related aspects of the test code such as coupling, cohesion, complexity, comprehensibility, maintainability, and change and defect proneness of the test code.

Nevertheless, the reported detection tools rely on imprecise detection rules (the number of method invocations and assertion statements in a unit test) due to the ambiguous definitions of the Eager Test smell.
This leads to inconsistent and questionable detection results~\cite{panichella2022test, Panichella2020Revisiting}. 
Consequently, the assessments of the prevalence and impact of the Eager Test smell in practice may have been inaccurate.
Meanwhile, practitioners highly agree that a good test case should be atomic, i.e., one test case should be testing one aspect of a requirement or one functionality~\cite{Kochhar2019Practitioners}, which is the exact opposite of eager tests.
Therefore, it is important to avoid, detect, and resolve this test smell.
However, they have expressed concerns about how the Eager Test smell has been defined and detected~\cite{Spadini2020Investigating,Schvarcbacher2019Investigating}.

In this study, we first reassess the definitions and detection rules of the Eager Test smell through a literature review.
Second, we propose a new, unambiguous definition of the Eager Test smell based on the definitions of the test smell found in the literature.
Third, we provide a novel detection heuristic (with pseudocode) that captures the proposed definition to overcome the weaknesses of existing detection rules and approaches.
Consistent with most discussions of test smells in the literature, the scope of the proposed heuristic is limited to unit tests and source code written in Java.
However, we also recognize the value of extending the Eager Test smell definition to non-object-oriented programming languages, such as C, to broaden the applicability of our approach across diverse programming paradigms.
Lastly, to assess the proposed heuristic, we apply it to a set of 300 test cases to detect eager tests.
We compare the results of the proposed heuristic with the existing detection rules.
Our assessment identify a high level of disagreement between the detection rules and highlights common patterns of eager tests and non-eager tests that can be detected by our heuristic but not by the existing rules.


The remainder of the paper is structured as follows: 
Section~\ref{sec:background} presents the basic concepts in unit testing and the related work.
Section~\ref{sec:methodology} details the design and procedure of the systematic literature review, the evaluation of the proposed heuristic, and the primary threats to validity.
Section~\ref{sec:resultAnalysis_RQ1_3},  Section~\ref{sec:resultAnalysis_RQ4}, and Section~\ref{sec:resultAnalysis_RQ5} present the results and analysis for each of the research questions posed in the study.
In Section~\ref{sec:discussion}, we discuss the selected findings related to the differences between our heuristic and the selected rules in detecting eager tests.
Finally, Section~\ref{sec:conclusion} concludes our paper.

\section{Background and related work} \label{sec:background}
In this section, we summarize the background notions with respect to unit testing, as well as the main related work discussing (1) test smells, (2) the quality of test cases and test suites, and (3) issues in the existing approaches for detecting eager tests.

\subsection{Unit testing}\label{sec:background_unitTesting}
The primary objective of unit testing is to ``enable the sustainable growth of a software project''~\cite{khorikov2020unit}.
While the scope of a test generally covers a collection of executable software components, the \textit{\textbf{scope} of a unit test} is focused on a relatively small executable~\cite{binder2000testing}.
In object-oriented programming, the smallest executable unit is a class object.
However, since test messages are sent to a method, it is actually a method scope testing~\cite{binder2000testing}.
Accordingly, we refer to the method being tested as the \textit{method under test} (MUT).
The class containing the MUT is called the \textit{class under test} (CUT), and the instance of this class, created in the test case, is termed the \textit{object under test} (OUT).
In this study, we adopt this terminology when defining the Eager Test smell and describing our heuristic for detecting it.

As with many terms in software development, the definition of a unit test comes with several nuances.
However, despite the variations, a unit test should consistently include three key elements as follows~\cite{khorikov2020unit, fowler2014unit}:
\begin{itemize}
    \item Verifying a small piece of code;
    \item Running significantly faster than other kinds of tests;
    \item Running in an isolated manner.
\end{itemize}

While the first two attributes are generally accepted, the third—what constitutes testing in an \textit{isolated manner}—often sparks debate in both practice and academia.
This issue of isolation forms the basis of the distinction between the \textit{classical}~\cite{Beck2003Test} and \textit{London} schools of unit testing~\cite{freeman2009growing}, also known as the Detroit and mockist schools, respectively.
The London school advocates for isolating the unit under test from its collaborators.
A \textbf{test unit}, in this case, is a \textit{unit of code}, which typically refers to a \textit{class}.
If a class depends on other classes, all such dependencies should be replaced with test doubles~\cite{meszaros2007xunit}.
In contrast, the classical approach focuses on ensuring that unit tests themselves are isolated from one another.
In this approach, a \textbf{test unit} is a \textit{unit of behavior}, which can be implemented by one or many classes.

Besides the three common elements, a unit test case also follows a common pattern, which is called the setup-stimulate-verify-teardown (S-S-V-T) cycle, as follows:
\begin{itemize}
    \item Acquire the necessary resources,
    \item Send one or more stimuli to the unit under test,
    \item Verify that the unit responds properly, and
    \item Release the acquired resources.
\end{itemize}

The importance of testing one method at a time in unit testing has been emphasized in the literature~\cite{khorikov2020unit, hamill2004unit, Tudose2021JUnit}.
There are several reasons for this approach.
Such tests are easier to understand and remain unaffected when other parts of the code change.
Additionally, they allow practitioners to refactor production code with confidence, as these tests provide clear messages that precisely identify the source of any breakage introduced by refactoring.
This recommendation aligns with what has been advised by researchers in order to avoid the Eager Test smell, which we explain further in detail in Section~\ref{sec:resultAnalysis_RQ1}.

\subsection{Test smells}
The concept of test smells originates from Fowler’s book on code smells~\cite{fowler2018refactoring}, where he explained that while code smells may not represent actual faults, they can pose quality issues that hinder maintenance or lead to bugs during later stages of development~\cite{tufano2015and}.
This idea was later expanded to test code by Deursen et al.~\cite{van2001refactoring}, who, in the context of unit testing for eXtreme Programming, described test smells as signs of problems in test code. Meszaros~\cite{meszaros2007xunit}, building on the work of Van Deursen et al., further developed the concept by identifying additional test smells specific to the xUnit testing framework. The most recent secondary study on test smell definitions was carried out by Garousi et al.~\cite{garousi2018smells}, which compiled a list of 139 test smells and provided an overview of approaches and tools to address them. Additionally, Aljedaani et al.~\cite{Aljedaani2021Test} offered a comprehensive review of 22 test smell detection tools developed by the research community.

\subsection{Quality of test cases and test suites}
After summarizing the main related work on test smells in the previous section, we now turn to the quality of test cases and test suites.
Test smells, as indicators of potential issues in test design, directly impact the overall quality of test cases and test suites. 
Hence, it is also essential to present the related work on the quality of test cases and test suites.

One of the pioneering studies in this area was conducted by Goodenough and Gerhard~\cite{goodenough1975toward}, who garnered significant attention from the research community by exploring test adequacy and their proposed definition of reliable test data.
Years later, Zhu et al.~\cite{zhu1997software} performed a comprehensive literature review on test adequacy criteria, which inspired further research into the structural quality of tests within various application domains and programming paradigms~\cite{kapfhammer2003family,lemos2007control,pei2019deepxplore}.
Additionally, some researchers began focusing on adapting software quality models to assess test case and test suite quality~\cite{neukirchen2008approach,athanasiou2014test}.
More recently, there has been an increased emphasis on incorporating feedback from practitioners to develop effective quality models for test cases and test suites~\cite{bowes2017how,grano2020pinsa,tran2019test}.

A recent literature review, conducted by the authors of this study, offers a comprehensive overview of the current state of test artifact quality in software testing~\cite{tran2021assessing}.
This review introduces a quality model for test cases and test suites, containing 30 quality attributes, with measurement details provided for 11 of these attributes.
The model also incorporates quality attributes adapted from ISO/IEC 25010:2011, offering a broader perspective on test case and test suite quality.
Additionally, the review identifies 11 context dimensions used to characterize the software-testing environments in which quality has been evaluated.
This quality model could prove valuable in developing guidelines or templates for designing test cases and test suites, as well as assessment tools for evaluating existing test artifacts.

\subsection{Issues in the existing approaches for detecting eager tests} \label{sec:background_issuesWithCurrentApproaches}
Panichella et al.~\cite{panichella2022test, Panichella2020Revisiting} conducted a manual analysis of 200 test suites generated by EVOSUITE and JTEXPERT and 49 manually written test suites to identify six types of test smells, including Eager Test.
The authors used the outcome of their manual analysis to assess the accuracy of the two detection tools~\cite{Bavota2012empirical, Peruma2020TsDetect} and to examine how the detected test smells reflect real problems in practice.
With respect to the Eager Test smell, the authors concluded that there is a need for a more precise approach to detecting the test smell as the current detection tools rely on highly inaccurate rules.

Panichella et al.~\cite{panichella2022test, Panichella2020Revisiting} also proposed a new criterion called `semantic coherence' to characterize non-eager tests.
More precisely, they stated that ``a test is semantically coherent if it asserts only (transitive) properties and attributes of the class under test that all relate to a single testing scenario.''
Their approach to identifying tested scenario(s) in a test case is to compare the asserted properties with the test's purpose, via its nomenclature and comments.
Nonetheless, it is unclear what the authors mean by asserted properties and how feasible it is to identify a test's purpose via its nomenclature and comments.
Hence, while we agree with the authors that a non-eager test should focus on only one testing scenario, the main limitation is that their approach to identifying the scenario depends on the readability of the code (how understandable the function names and comments are).
Also, it seems to require subjective assessment that would require human involvement and perhaps would be difficult to automate.

Compared to Panichella et al.~\cite{panichella2022test, Panichella2020Revisiting}, we propose an unambiguous definition of the Eager Test smell and a concrete heuristic (with pseudocode) that operationalizes the definition and can be deterministically applied to detect eager tests.
Furthermore, our heuristic works at the test case level, which provides a better understanding of the presence of eager tests than the annotation at the test suite level by Panichella et al.
They also acknowledged that their analysis could lead to overestimating the presence of test smells at a test case level.

The second related work that we want to discuss is from Pizzini et al.~\cite{Pizzini2022Automatic}. They summarized different definitions of the Eager Test smell and studied the existing detection approaches.
They argued that the existing approaches are not good enough to capture eager tests as they produce a high number of false positives, i.e., wrongly label tests as having the Eager Test smell.
Based on this argument, they proposed a new solution for identifying the test smell.
They further implemented a refactoring method to remove eager tests in practice automatically.
One of the potential threats to their study's validity is that their study of the definitions and existing detection approaches was not systematic.
For this reason, their study contained an imprecise reference to the original definitions and an incomplete list of existing approaches.

Compared to the studies of Panichella et al.~\cite{panichella2022test, Panichella2020Revisiting} and Pizzini et al.~\cite{Pizzini2022Automatic}, we argue that our study provides a more thorough review of how the Eager Test smell has been defined and detected.
In addition, our manual analysis showed that our heuristic could provide a better detection accuracy than the existing detection approaches when assessing eager tests at a fine-grained level, i.e., the test case level.

\section{Methodology}\label{sec:methodology}
The goals of this study are (1) to understand how the Eager Test smell has been defined and detected in the literature and (2) to address the limitations of existing rules and approaches to detecting eager tests.
To achieve these goals, our study aims to answer the following research questions:
\begin{itemize}
    \item RQ1. How has the Eager Test smell been defined in the literature?
    \item RQ2. What rules and approaches have been proposed in the literature to detect eager tests?
    \item RQ3. What are the limitations of the existing detection rules and approaches?
    \item RQ4. How can the identified limitations be addressed?
    \item RQ5. To what degree does the level of recognized eager tests change when these limitations are addressed?
    
\end{itemize}

We conducted a literature review to answer RQ1, RQ2, and RQ3.
With the review, we wanted to understand how the Eager Test smell was originally defined and any shortcomings of the current detection rules and approaches.
To answer RQ4, we proposed a novel, more precise definition of the Eager Test smell and constructed a heuristic that can operationalize the proposed definition in order to detect eager tests for unit test cases with source code written in Java.
The proposed definition and heuristic were built based on the information obtained by RQ1-3.
Finally, we addressed RQ5 by performing a manual eager test classification of 300 unit test cases (both manually written and automatically generated) written in Java.
The assessment was based on the proposed heuristic and the existing detection rules and approaches identified by RQ2.

\subsection{Literature review -- Search process and data extraction}
The literature review on definitions and detection rules of the Eager Test smell was conducted through two searches.
First, we looked for existing studies that had synthesized definitions of the Eager Test smell.
This first search was performed on March 17, 2023, by searching for reviews on test smells on Scopus using the search string ``(TITLE-ABS-KEY(``review'' OR ``map'' OR ``mapping'') AND TITLE-ABS-KEY (``test smells''))''.
The first search returned 15 papers.
By scanning the abstracts, we (three of the co-authors) selected six studies that are literature reviews on test smells.
The three co-authors read the full text of these six papers to check if the papers present definitions of the Eager Test smell they have found or synthesized.
We found two such studies~\cite{garousi2018smells,rwemalika2023smells}.
From these two studies, we found 28 papers (cited in~\cite{garousi2018smells}) and 15 papers (cited in~\cite{rwemalika2023smells}) that included the definitions of the Eager Test smell.

Since the first relevant review~\cite{garousi2018smells} was not recent (2016), while the second relevant review~\cite{rwemalika2023smells} focused on a specific type of test cases (system user interactive tests), we decided to perform a second search to identify more recent work.
Although we were interested in Eager Test only, for the search, we used the more general keyword of ``test smell'' instead to be more inclusive.
The second search was conducted on Scopus using this search string: ``TITLE-ABS-KEY (``test smell'') AND (LIMIT-TO (SUBJAREA, ``COMP'')) AND (LIMIT-TO (DOCTYPE, ``cp'') OR LIMIT-TO (DOCTYPE , ``ar''))''.
The search returned 93 papers.
In total, we collected 136 papers (93 from the second search + 28 from the first relevant review~\cite{garousi2018smells} + 15 from the second relevant review~\cite{rwemalika2023smells}).
After removing the duplicates, non-peer-reviewed literature (based on expert opinion), and studies that only mentioned the test smells without descriptions, we had a final list of 56 papers (available in our supplementary document, which can be accessed via the following link: \url{https://figshare.com/s/34c20027fb4e5c2344d6}) for data extraction.

During data extraction, we collected information about both the Eager Test and Assertion Roulette test smells, as Assertion Roulette was reported as identical to the Eager Test smell by Garousi and K{\"u}{\c{c}}{\"u}k~\cite{garousi2018smells}.
The following information was extracted from the selected 56 papers: (1) definition or description of the test smell (Eager Test or Assertion Roulette), (2) source of definition (references to other papers or the authors' interpretation), (3) detection rules, (4) source of detection rules (references to other papers or the authors' interpretation). 
All authors reviewed the data extraction form.
After that, the first author completed the extraction process.
One co-author reviewed the extraction result of 10\% of the selected papers (randomly selected).
Since there was a 100\% agreement among two co-authors, we decided not to pursue further validation of the extraction.

\subsection{Evaluation of proposed heuristic}
To evaluate our heuristic for detecting eager tests, we manually assessed a set of 300 test cases, both manually written and automatically generated, using the heuristic and the existing detection rules found in the literature.
To compare the level of agreement in the eager test classification outcomes among our heuristic and the detection rules, we used Cohen's kappa, which can be used to measure the degree of agreement in nominal scales between two raters~\cite{cohen1960coefficient}.
We then followed Landis and Kock's guideline~\cite{landis1977measurement} to interpret the kappa values.
The raw data (the number of test cases classified as eager tests or not by each detection rule and the heuristic) is included in Appendix~\ref{appendix:ET_classification_outcome}.
Note that we call our heuristic \textit{EagerID} in Appendix~\ref{appendix:ET_classification_outcome} for the sake of simplification.

\subsubsection{Test case selection}\label{sec:testCaseSelection}
To compare our heuristic with existing eager test detection rules, we manually analyzed 300 Java unit test cases used in recent studies by Panichella et al.\cite{panichella2022test} and Sharma et al.\cite{tushar2023investigating}.
These studies were selected because they focus on assessing eager tests in unit testing, aligning with our goals, and their selection of test cases was well-motivated and from well-known sources of open-source software systems.

In particular, these 300 test cases came from two widely-used datasets: the SF110 dataset~\cite{fraser2014large} (containing 110 projects from SourceForge.net) and the RepoReapers dataset~\cite{munaiah2017curating} (generated from 1,857,423 GitHub repositories).
Specifically, among the 300 selected test cases, 100 automatically generated test cases came from 17 different projects (from the SF110 dataset), while 200 manually written tests were from eight different projects (from the SF110 and the RepoReapers datasets).

Another reason for using the same test cases as Panichella et al. and Sharma et al. is that they shared their manual assessment of eager tests.
Hence, by analyzing the same data, we could cross-validate our results with theirs, as explained in detail below.
It is also worth noting that we did not assess all test cases analyzed in these two studies, as our heuristic requires a clear understanding of the production code, which can consume a considerable amount of time.
Given our study's limited human resources, we randomly selected 300 test cases for our manual classification.


In Sharma et al.'s study, the authors used one of the common detection rules found in the literature (``a test method calls multiple production methods") to manually classify 427 test cases (manually written).
Note that the authors did not specify the threshold used in their manual classification and whether they counted any method calls from the production code or from the class under test only.
However, based on the test cases classified as eager tests in their published assessment data, we could clearly see that the counted method calls were from the production code, and the number of such calls was always more than four.
With this information, we argued that their detection rule was identical to the detection rule DR2.3 in Table~\ref{tab:comparison_detectionRules}.
As their manual assessment is accessible, we were able to compare our eager test classification with theirs and explored the similarities and differences in interpreting the same detection rule.

With Panichella et al.'s study, the authors used their own proposal (``At least 2 assertions in a test case and at least one
assertion is not on the result of a get met" - DR4 in Table~\ref{tab:comparison_detectionRules}) to manually identify eager tests in their set of 149 test suites (100 automatically generated + 49 manually written ones).
Their assessment was conducted at the test suite level, while we assessed eager tests at a finer-grain level, i.e., the test case level.
For this reason, we could not study the similarities and differences in interpreting the detection rule as we did with Sharma et al.'s study.
Nevertheless, one of their findings drew our attention.
The authors concluded that many test suites were classified as eager according to their detection rule DR4 (20\% of the automatically generated test suites and 80\% of the manually written test suites).
Since we wanted to verify if we could observe the same pattern with our heuristic at the test case level, it was still worth it for us to work on the same data set.

\subsection{Validity threats}\label{sec:validityThreats}
In the following, we discuss potential threats to the validity of the literature review, and the heuristic and its validation.
\subsubsection{Threats to construct validity}
One of the main challenges in conducting this study was related to the literature review.
We followed the standard procedure when conducting the review.
Nonetheless, there were two steps, including the search validation, and the quality assessment of the selected studies that we did not perform, which might affect the review's outcome.
Also, since our search was on Scopus only, we might have missed other relevant studies that could be found via other search engines.
We mitigated this problem by including studies cited by two recent literature reviews on eager tests to expand our selected studies collection.
Other common threats to the validity of secondary studies are bias in paper selection and data extraction.
To address these problems, the paper selection result and the data extraction form were reviewed and discussed by all authors.
Also, two authors extracted data for a subset of seven papers.
There was a high consistency in the data extracted by the two authors, and the final data extraction result was also reviewed by all authors.

Another threat was related to how we understood the original definitions of the Eager Test smell.
As mentioned earlier, the definitions are rather informal, which gives room for interpretation.
Therefore, there could be details or ideas in the definitions that we might not interpret the same way as the authors of the definitions.
However, we argue that our heuristic captures the essence of the Eager Test smell as we thoroughly studied the original definitions.
On top of that, we presented concrete patterns of both eager and non-eager test cases that could be highlighted by our heuristic only and not by the other detection rules in the literature.

Connected to the threat above, another challenge was our interpretation of the detection rules.
To make sure that we understood the detection rules sufficiently, we looked at both the original references and the source code of the tools that implemented the detection rules.
Nevertheless, there could still be details that we missed, which can lead to a different understanding, as shown by the different assessment results between us and Sharma et al. regarding the detection rule DR2.3 (details in Section~\ref{sec:comparisonWithPanichellaAndSharma}).

Another threat was how we conducted our manual assessment.
Even though the assessment was mainly done by the first author, the outcomes were discussed in iterations with two other co-authors via face-to-face discussions and workshops.
Also, for each assessed test case in our manual assessment, we attached the production code, from which we collected the necessary information to execute our heuristic.
As such, we are confident that our assessment based on the heuristic is highly traceable and internally consistent.

\subsubsection{Threats to external validity}
In this study, we assessed the presence of eager tests in a set of 300 test cases.
The choice of the tests used in this study might potentially impact the generalizability of our findings.
As the sample set was rather small, we might potentially miss different root causes of the differences between our heuristic and the existing detection rules.
Nonetheless, our specific findings regarding what our heuristic can contribute above the existing rules remain valid despite the small sample set of test cases.


\section{Eager Test definitions, detection rules and limitations (RQ1--RQ3)}\label{sec:resultAnalysis_RQ1_3}
\subsection{Eager Test definitions (RQ1)}\label{sec:resultAnalysis_RQ1}
\subsubsection{Sources of definitions}
The first important finding from the literature review was that there are actually two primary sources of definitions for Eager Test and Assertion Roulette test smells: van Deursen et al.~\cite{van2001refactoring} and Meszaros~\cite{meszaros2007xunit} (Table~\ref{tab:testSmellsDefintions}).
However, not all studies cited these two primary sources for their test smells' definitions.
Among the 56 selected studies, 47 studies provided definitions of the Eager Test smell, but only 38 out of the 47 studies cited the primary sources.
The other 9 ($47-38$) studies either did not mention the source of their test smells' definitions or cited different studies instead.
Similarly, among 48 studies that provided definitions of the Assertion Roulette smell, only 41 studies referred to the primary sources.
It is also worth noting that the definitions given by van Deursen et al.~\cite{van2001refactoring} were more often cited than the definitions given by Meszaros~\cite{meszaros2007xunit} (as shown in Table~\ref{tab:testSmellsDefintions}).

Between the two primary sources, van Deursen et al.~\cite{van2001refactoring} were the first ones who introduced the concept of test smells to convey poorly designed unit tests.
Meszaros~\cite{meszaros2007xunit} acknowledged the introduced test smells and offered a broader point of view on the same topic by classifying test smells into three levels: project, behavior, and code-level smells.
A test smell at a higher level contains lower-level test smells.
A common factor between the two sources of definitions is that the authors defined the test smells rather informally, giving room for interpretation, which is further discussed in the subsequent section.

With respect to the association between Eager Test and Assertion Roulette, Meszaros noted that Eager Test can be the cause of Assertion Roulette.
However, neither Meszaros nor van Deursen et al. discussed their equivalence.
For the Eager Test smell, its definitions from both sources focus on either the number of methods or functionalities to be tested.
Meanwhile, for the Assertion Roulette smell, both sources emphasize the problem with assertions having no explanation to assist the test failure analysis.
The claimed equivalence was found only in a later study~\cite{garousi2018smells}, where the authors interpreted the original definitions.

\begin{table}
{    
    \footnotesize
    \begin{center}
    \caption{Definitions of Eager Test and Assertion Roulette given by van Deursen et al.~\cite{van2001refactoring} and Meszaros~\cite{meszaros2007xunit}}
    \label{tab:testSmellsDefintions}    
    \begin{threeparttable}    
    \begin{tabular}{L{0.14\linewidth}p{0.4\linewidth}p{0.3\linewidth}}
        \toprule
        \textbf{Test smell} & \textbf{van Deursen et al.}~\cite{van2001refactoring} & \textbf{Meszaros}~\cite{meszaros2007xunit} \\
        \midrule
        Eager Test\tnote{1} & When a test method checks several methods of the object to be tested, it is hard to read and understand, and therefore more difficult to use as documentation. Moreover, it makes tests more dependent on each other and harder to maintain. (cited by 32 studies) & The test verifies too much functionality in a single test method. (cited by 6 studies) \\ [4pt]
        Assertion Roulette\tnote{2} & “Guess what’s wrong?” This smell comes from having a number of assertions in a test method that have no explanation. If one of the assertions fails, you do not know which one it is. (cited by 35 studies) & It is hard to tell which of several assertions within the same test method caused a test failure. (cited by 6 studies) \\
        \bottomrule
    \end{tabular}
    \begin{tablenotes}
        \footnotesize
        \noindent
        \begin{minipage}[c]{1\linewidth}
            \item[1] The definition was provided in 47 studies.
            \item[2] The definition was provided in 48 studies.
        \end{minipage} 
    \end{tablenotes}
    \end{threeparttable}
    \end{center}
}
\end{table}

\subsubsection{Interpretations of the original definitions}
Since the definitions of the Eager Test and Assertion Roulette smells given by van Deursen et al.~\cite{van2001refactoring} and Meszaros~\cite{meszaros2007xunit} are rather ambiguous, researchers studying these two types of test smells provided their own interpretations.
Table~\ref{tab:eagerTest_interpretedDefinitions} and Table~\ref{tab:assertionRoulette_interpretedDefinitions} categorize those interpreted definitions of Eager Test and Assertion Roulette, respectively.
By comparing the two tables, we can see the overlap in interpreted definitions between Eager Test and Assertion Roulette.
More specifically, a common definition, namely \textit{having multiple assertions}, was used to characterize both types of test smells in several studies~\cite{garousi2018smells, Tahir2016Empirical, Camara2021On}.

When looking more closely at each type of test smell, we can see that the Eager Test has more definition variations than the Assertion Roulette.
With Assertion Roulette, the interpreted definitions mainly revolve around one aspect: assertion(s) without explanation.
For Eager Test, the interpreted definitions scatter from a unit test having multiple method calls of the object or class under test to a unit test checking multiple methods or functionalities.
Note that having multiple method calls is not the same as checking/verifying several methods in a test case.
Ultimately, the findings show diverse interpretations of the Eager Test smell in the literature, even though they are mainly based on the same two definitions (see Table~\ref{tab:testSmellsDefintions}).

\begin{table}
{    
    \footnotesize
    \begin{center}
    \caption{Interpreted definitions of \textbf{Eager Test}}
    \label{tab:eagerTest_interpretedDefinitions}
    \begin{threeparttable}    
    \begin{tabular}{p{0.6\linewidth}p{0.08\linewidth}p{0.08\linewidth}p{0.035\linewidth}}
        \toprule
        \textbf{Interpreted definition} & \multicolumn{3}{c}{\textbf{Number of studies}\tnote{*}}  \\[4pt]
        & \textbf{Based on DS1}\tnote{1} & \textbf{Based on DS2}\tnote{2} & \textbf{NA}\tnote{3} \\[4pt]
        \midrule
        Checking multiple methods of the object under test & 11 & 1 &  1\\
        Checking multiple methods of the class under test & 5 & 1 & 0\\
        Checking or using multiple methods of the class under test & 2 & 0 & 0 \\
        Checking multiple different functionalities & 2 & 2 & 3  \\
        Having multiple methods of the object under test & 4 & 0 & 1  \\
        Having multiple methods of the class under test	& 2 & 0 & 1  \\
        Having multiple calls to multiple production methods & 4 & 2 & 2  \\
        Exercising multiple non-related features of the system under test & 1 & 0 & 0 \\
        Having multiple assertions & 0 & 0 & 1 \\
        \bottomrule
    \end{tabular}
    \begin{tablenotes}
        \footnotesize
        \noindent
        \begin{minipage}[c]{1\linewidth}
            \item[*] The studies' references are available via \\
            \url{https://figshare.com/s/34c20027fb4e5c2344d6} 
            \item[1] DS1 - Based on the definition of Eager Test given by van Deursen et al.~\cite{van2001refactoring}
            \item[2] DS2 - Based on the definition of Eager Test given by Meszaros~\cite{meszaros2007xunit}
            \item[3] NA - The authors did not specify the source(s) of definition
        \end{minipage} 
    \end{tablenotes}
    \end{threeparttable}
    \end{center}
}
\end{table}

\begin{table}
{    
    \footnotesize
    \begin{center}
    \caption{Interpreted definitions of \textbf{Assertion Roulette}}
    \label{tab:assertionRoulette_interpretedDefinitions}
    \begin{threeparttable}
    \begin{tabular}{p{0.6\linewidth}p{0.08\linewidth}p{0.08\linewidth}p{0.035\linewidth}}
        \toprule
        \textbf{Interpreted definition} & \multicolumn{3}{c}{\textbf{Number of studies}\tnote{*}} \\[4pt]
        & \textbf{Based on DS1}\tnote{1} & \textbf{Based on DS2}\tnote{2} & \textbf{NA}\tnote{3} \\[4pt]
        \midrule
        Having multiple assertion statements without explanation & 27 & 3 & 6 \\
        Having multiple assertion statements that do not provide any description of why they failed & 2 & 0 & 1 \\
        Having multiple assertion statements with at least one	that does not provide any description of its failure & 3 & 0 & 0\\
        Having multiple assertions	& 2 & 0 & 0 \\
        Having an assertion statement that does not contain an explanation & 0 & 0 & 1 \\
        Being hard to tell which of several assertions failed & 1 & 3 & 1 \\
        \bottomrule
    \end{tabular}
    \begin{tablenotes}
        \footnotesize
        \noindent
        \begin{minipage}[c]{1\linewidth}
            \item[*] The studies' references are available via \\
            \url{https://figshare.com/s/34c20027fb4e5c2344d6} 
            \item[1] DS1 - Based on the definition of Eager Test given by van Deursen et al.~\cite{van2001refactoring}
            \item[2] DS2 - Based on the definition of Eager Test given by Meszaros~\cite{meszaros2007xunit}
            \item[3] NA - The authors did not specify the source(s) of definition
        \end{minipage} 
    \end{tablenotes}
    \end{threeparttable}
    \end{center}
}
\end{table}

\subsection{Eager Test detection rules (RQ2)}\label{sec:resultAnalysis_RQ2}
Since each type of test smells, especially Eager Test, has been interpreted differently, their detection rules are also divergent, as summarized in Table~\ref{tab:eagerTest_detectionRules} and Table~\ref{tab:assertionRoulette_detectionRules}.
On the one hand, with Assertion Roulette, most of the proposed detection rules are about the assertion statements having no explanation except one study~\cite{junior2020survey}, which stated that an Assertion Roulette test is the case of ``putting together tests that could be run separately.''
This particular detection rule of Assertion Roulette is broad enough to cover the Eager Test smell as well.
On the other hand, while the proposed detection rules for Eager Test are more divergent, they can be grouped into three main categories: (1) the number of invoked methods, (2) the number of cycles of non-verification instructions followed by verification instructions, (3) the textual similarity among the tested methods.
The first category of detection rules is the major one (12 out of 14 studies), while the other two categories each came from two other studies.

It is worth reemphasizing that the reason for including Assertion Roulette in the review was because Assertion Roulette and Eager Test were claimed to be the same~\cite{garousi2018smells} and we did not want to miss finding definitions or detection rules of the Eager Test smell.
Nevertheless, the analysis presented earlier shows that the original definition of Assertion Roulette is very different from Eager Test, and the overlap in the interpreted definitions and rules does not cover more relevant aspects than those targeted explicitly for Eager Test.
Therefore, our following discussion focuses on the definitions and detection rules for Eager Test only (as presented in Tables~\ref{tab:testSmellsDefintions}, \ref{tab:eagerTest_interpretedDefinitions}, and~\ref{tab:eagerTest_detectionRules}).

\begin{table}
{    
    \footnotesize
    \begin{center}
    \caption{Proposed detection rules for \textbf{Eager Test}}
    \label{tab:eagerTest_detectionRules}
    \begin{threeparttable}
    \begin{tabular}{p{0.55\linewidth}p{0.15\linewidth}p{0.15\linewidth}}
        \toprule
        \textbf{Proposed rule} & \textbf{Source of definition}\tnote{1} & \textbf{Number of studies}\tnote{2} \\
        \midrule
         Number of production type method invocations in a test case & S1, S2 & 7 \\
         JUnit classes having at least one method that uses more than one method of the tested class & S1 & 4 \\
         More than one method call from class under test in a test case & S1 & 1 \\
         The test method instructions contain at least two cycles of non-verification instructions followed by verification instructions & S1 & 1 \\
         The low textual similarity among the method calls in a test case & S1 & 1 \\
         The test must have more than one assertion and at least one assertion is not on the result of a get method & S1 & 1 \\
        \bottomrule
    \end{tabular}
    \begin{tablenotes}
        \footnotesize
        \noindent
        \begin{minipage}[c]{1\linewidth}
            \item[1] S1 - Based on the definition of Eager Test given by van Deursen et al.~\cite{van2001refactoring}  \\
            S2 - Based on the definition of Eager Test given by Meszaros~\cite{meszaros2007xunit} \\
            NA - The authors did not specify the source(s)
            \item[2] The studies' references are available via \\
            \url{https://figshare.com/s/34c20027fb4e5c2344d6} 
        \end{minipage} 
    \end{tablenotes}
    \end{threeparttable}
    \end{center}
}
\end{table}

\begin{table}
{    
    \footnotesize
    \begin{center}
    \caption{Proposed detection rules for \textbf{Assertion Roulette}}
    \label{tab:assertionRoulette_detectionRules}
    \begin{threeparttable}
    \begin{tabular}{p{0.5\linewidth}p{0.15\linewidth}p{0.17\linewidth}}
        \toprule
        \textbf{Proposed rule} & \textbf{Source of definition}\tnote{1} & \textbf{Number of studies}\tnote{2} \\
        \midrule
         Some tests fail and it is not possible to identify the failure cause & S1, S2 & 1 \\
         Putting together tests that could be run separately & S1, S2 & 1 \\
         The number assertions without description & S1, NA & 13 (3 from NA)\\
         The number of assertions without additional/optional argument & S1 & 2\\
        \bottomrule
    \end{tabular}
    \begin{tablenotes}
        \footnotesize
        \noindent
        \begin{minipage}[c]{1\linewidth}
            \item[1] S1 - Based on the definition of Eager Test given by van Deursen et al.~\cite{van2001refactoring}  \\
            S2 - Based on the definition of Eager Test given by Meszaros~\cite{meszaros2007xunit} \\
            NA - The authors did not specify the source(s)
            \item[2] The studies' references are available via \\
            \url{https://figshare.com/s/34c20027fb4e5c2344d6} 
        \end{minipage} 
    \end{tablenotes}
    \end{threeparttable}
    \end{center}
}
\end{table}

\subsection{Limitations of the Eager Test detection rules (RQ3)}\label{sec:resultAnalysis_RQ3}
At first sight, the definitions of the Eager Test smell given by van Deursen et al.~\cite{van2001refactoring} and Meszarous~\cite{meszaros2007xunit} seem quite different as the former focused on methods of the object under test while the latter stressed the functionalities under test.
To further investigate the difference, we looked into the extra details provided by Meszarous for the Eager Test smell.
Meszarous explained that the root cause of the Eager Test smell is that practitioners verify multiple ``test conditions'' in a single test.
The author further described a test condition as follows:

``A test condition is a particular behavior of the system under test (SUT) that we need to verify. It can be described as the following collection of points:

\parbox{\textwidth}
    {
        If the SUT is in some state \textbf{S1}, and \\
        I exercise the SUT in some way \textbf{X}, then \\
        the SUT should respond with \textbf{R} and \\
        the SUT should be in state \textbf{S2}.''
    } \\

In the context of unit testing, it is reasonable to consider \textbf{X} as a method under test.
Hence, this definition of the Eager Test smell given by Meszarous is also oriented around verifying multiple methods, which is ultimately aligned with the definition from van Deursen et al.~\cite{van2001refactoring}, i.e., ``a test method checks several methods of the object to be tested'' (details in Table~\ref{tab:testSmellsDefintions}).

Nevertheless, as we can see from the literature review, not all interpreted definitions of the Eager Test smell (Table~\ref{tab:eagerTest_interpretedDefinitions}) focus on the same issue as the original definitions.
The interpreted definitions mainly focused on the number of methods invocations, which is not the same as verifying multiple methods.
One or more invocations of methods other than the method under test might be required as part of the test setup or to acquire intermediate results in order to verify the outcome(s) of the method under test.
This issue with the interpreted definitions has also been raised by other researchers~\cite{panichella2022test,white2022TCTracer}.
Therefore, focusing on the number of method invocations does not adequately capture the original definitions of the Eager Test smell.
However, counting the number of method invocations has been used as a lodestar to establish most detection rules (Table~\ref{tab:eagerTest_detectionRules}) and verify the detection results.

Furthermore, our observation of the current state of the art is that many studies~\cite{breugelmans2008testq, Bavota2012empirical, Tufano2016empirical, Bleser2019Assessing, Palomba2018Automatic, Peruma2020TsDetect, Santana2022Refactoring} have used these imprecise interpreted definitions and detection rules to build their detection tools which are in turn used to further analyze the prevalence of eager tests and their impact on the quality of test code (in terms of Comprehensibility and Maintainability).
The problem is that this introduces false positives, as reported in the recent studies of Panichella et al.~\cite{panichella2022test, Panichella2020Revisiting}.

On top of that, the assessment is not agreed upon by practitioners~\cite{Spadini2020Investigating, Schvarcbacher2019Investigating} even though practitioners highly agree that a good test case should be atomic, i.e., one test case should be testing one aspect of a requirement or one functionality~\cite{Kochhar2019Practitioners}, which is the exact opposite of eager tests.
Hence, the current assessment of the prevalence and impact of eager tests requires re-evaluation as it might not be adequate.

\section{Consolidated Eager Test definition and heuristic (RQ4)}\label{sec:resultAnalysis_RQ4}
To address the issues with the existing detection rules, we propose a more precise definition of the Eager Test smell and a novel heuristic that can operationalize the proposed definition to detect eager tests. In this section, we describe how we derived our definition of the Eager Test smell from the original definitions given by van Deursen et al.~\cite{van2001refactoring} and Meszarous~\cite{meszaros2007xunit} and well as how we constructed our heuristic.

\subsection{Proposed definition of the Eager Test smell}
According to the original definition of the Eager Test smell given by van Deursen et al.~\cite{van2001refactoring} (``a test method checks several methods of the object to be tested''), it is unclear whether checking \textit{multiple method calls} of \textit{the same method} is acceptable, i.e., classified as not eager.
Likewise, Meszarous' definition emphasizes ``verifying multiple ``test conditions'' in a single test'', in which we argued that a ``test condition'' also means a method to be tested (more details in Section~\ref{sec:resultAnalysis_RQ3}).
Hence, with both of these broad definitions, it is unclear whether a test case is classified as an eager test when it verifies the outcomes of \textit{multiple method calls} of \textit{the same method}.

Nevertheless, we argue that it is important to be clear about the case of verifying multiple method calls of the same method.
For example, a test case can invoke multiple method calls of a single method to check corner cases (negative value, zero, positive value). 
Since the object under test (OUT) behaves differently in each corner case, each method call presents a different behavior of OUT.
In other words, the test case actually tests different behaviors of OUT even though there is only one method under test.
Hence, our interpretation of these definitions is that a test case is an eager test if it verifies the outcomes of multiple method calls, disregarding whether these calls are from the same method or not.
Based on our interpretation, our novel, more precise definition of the Eager Test smell is as follows:

\begin{framed}
\noindent A test case is NOT an eager test when all of its assertions assess the outcome(s) of a single method call of the class under test (CUT).
\end{framed}

The definition is based on the analysis of what a method does (presented by its outcome(s)) and what is actually assessed by assertion statements.
The outcome(s) of a method call can be: the return value/ object(s) AND/OR the state change of an object caused by the method call.
The object with the changed state can either be from the class under test (CUT) or from another class.
The second case (object from a class different from CUT) occurs when the object is passed by parameter to the method under test (MUT).

\subsection{Description of the heuristic}
In this section, we describe our heuristic which can operationalize our proposed definition of the Eager Test smell in the context of unit test cases written in Java.

\subsubsection{Method classification}\label{sec: methClassification}
Since our proposed definition of the Eager Test smell is based on the analysis of what a method does (presented by its outcome(s)), we need to have a systematic classification of method types.
In this study, we relied on existing taxonomies~\cite{dragan2006reverse, dragan2009using, moreno2012jstereocode} of method stereotypes in the Object-oriented (OO) paradigm.
These taxonomies categorize methods into four main types: creational, mutator, collaborator, and accessor.
Table~\ref{tab:meth_stereotype} presents these main types and their sub types.
The table also shows how we adapt the method stereotypes proposed by the taxonomies in our heuristic.

\begin{table*}[h]
{    
    \footnotesize
    \begin{center}
    \begin{adjustwidth}{-1cm}{}
    \caption{Method stereotypes as defined by Moreno and Marcus~\cite{moreno2012jstereocode} and our adaptations}
    \label{tab:meth_stereotype}
    \begin{tabular}{p{0.07\textwidth} p{0.07\textwidth}p{0.15\textwidth}p{0.1\textwidth}p{0.28\textwidth}p{0.28\textwidth}}
    \toprule
    \textbf{Category} & \textbf{Stereotype} & \textbf{Description \cite{moreno2012jstereocode}} & \textbf{Action} & \textbf{Reasons} & \textbf{Our description} \\
    \midrule
    
    Accessor & Get & Returns a local field directly & Included & 
    Needed for the heuristic & Returns an object's attribute without modifying the attribute. \\
    \hdashline
    
    Accessor & Predicate & Returns a Boolean value that is not a
    local field & Merged with Property & \multirow{2}{0.3\textwidth}{We are interested in only the difference between a \textbf{get} and the others (\textbf{predicate} and \textbf{property}). Therefore, we group \textbf{predicate} and \textbf{property} into one type, called \textbf{producer}} & \multirow{2}{0.3\textwidth}{A \textbf{producer} computes a result based on an object's attribute(s). It means that a \textbf{producer} does not change the object's state, and its return type is not of interest. A \textbf{producer} can return a Boolean result (as a \textbf{predicate}) or have other return types (as a \textbf{property}).}\\
    & & & & & \\    
    & & & & & \\
    Accessor & Property & Returns information about local fields & Merged with Predicate & &\\
    \hdashline
    
    Accessor & Void-accessor & Returns information about local fields through the parameters & Excluded & It does not make sense in the context of Java programming. & N/A \\
    \midrule
    
    Mutator & Set & Changes only one local field & Merged with Command and Non-void command & \multirow{3}{0.3\textwidth}{We are not interested in different stereotypes under this category as the number of attributes changed by a mutator and its return type is not needed for the heuristic.} & \multirow{3}{0.3\textwidth}{A \textbf{mutator} is a method that changes the state of the object to which it belongs, i.e., must change one or more attributes of the object and optionally have a return value/object.} \\
    Mutator & Command & Changes more than one local fields & Merged with Set and Non-void command & & \\
    Mutator & Non-void command & Command whose return type is not void or Boolean & Merged with Set and Command & & \\
    \midrule

    Creational & Constructor & Invoked when creating an object & Merged with Copy-constructor and Factory & \multirow{3}{0.3\textwidth}{How an object is created is not relevant to the heuristic. Hence, we merged Constructor, Copy-constructor and Factory into one called \textbf{Creational}} & \multirow{3}{0.3\textwidth}{A \textbf{creational} method is responsible for creating an object.} \\
    Creational & Copy-constructor & Creates a new object as a copy of the existing one & Merged with Constructor and Factory & & \\
    Creational & Factory & Instantiates an object and returns it & Merged with Constructor and Copy-constructure & & \\
    \hdashline
    Creational & Destructor & Performs any necessary cleanups before the object is destroyed & Excluded & It does not make sense in the context of Java programming. & N/A \\
    \midrule

    Collabora-tional & Collaborator & Connects one object with another type of objects  & Excluded & \multirow{3}{0.3\textwidth}{A \textbf{Collaborational}, according to the authors, works on objects of classes different from itself and can also be either an accessor or mutator. Hence, the main difference is which object the method works on. Since this differentiation on the target object is irrelevant to our heuristic, we only care if a method is a mutator or an accessor instead of further classifying it as a collaborator-mutator or collaborator-accessor.} & \multirow{3}{0.3\textwidth}{N/A} \\
    & & & & & \\    
    & & & & & \\
    & & & & & \\    
    & & & & & \\
    Collabora-tional & Controller & Provides control logic by invoking only external methods & Excluded  & & \\
    Collabora-tional & Local controller & Provides control logic by invoking only local methods & Excluded  & & \\
    \bottomrule
    \end{tabular}
    \end{adjustwidth}
    \end{center}
}
\vspace{5mm}
\end{table*}

In addition to the adapted method types, we propose two sub types for a \textbf{producer}.
Note that these subtypes make sense only in the context of unit testing.
\begin{itemize}
    \item \textbf{Internal producer}: the producer method is implemented (or overridden) in the method's class, i.e., the class under test (CUT).
    \item \textbf{External producer}: the producer method is implemented in a class different from the class under test.
    This producer method is called to compute some result based on the object under test's attribute(s).
    For instance, this external class can be the parent class and any other classes in the external libraries.
\end{itemize}

Note that we do not distinguish \textit{internal} and \textit{external} for \textbf{get}, \textbf{creational}, and \textbf{mutator} methods.
Since a \textbf{get} method is simply used to retrieve an object's attribute(s), it does not matter whether the get method is implemented in the CUT or not.
Likewise, if a method like \textbf{creational} and \textbf{mutator} is implemented outside CUT, there is no reason to test this method in CUT.
Testing such a method should be handled by a separate test case at the class the method belongs to so that there is no mix of the two units (CUT) in one single test case.
Hence, \textbf{creational} and \textbf{mutator} methods we refer to in this heuristic are the methods implemented in CUT.
Table~\ref{tab:methodTypes} presents the final list of method types used in our heuristic.

\begin{table*}
\caption{Method types used in our heuristic}
\label{tab:methodTypes}
\begin{threeparttable}
\begin{tabular}{p{0.1\textwidth}p{0.12\textwidth}p{0.05\textwidth}p{0.1\textwidth}p{0.1\textwidth}p{0.1\textwidth}p{0.1\textwidth}p{0.12\textwidth}}
\hline
\rowcolor[HTML]{EDEDED}
Method Type & Sub type& CUT\tnote{*}&     \multicolumn{5}{c}{Effect}\\
\hline
 & & & Initialize attribute& Retrieve attribute& Change attribute& Return attribute&Return value/object\\
\midrule
Creational        &   & \checkmark &     \checkmark&-&-&-&[Optional]\\
\midrule
Mutator           &   & \checkmark &     -&\checkmark &\checkmark &-&[Optional]\\
\midrule
Accessor&   Get               &   \checkmark\tnote{**} &                           -&\checkmark &-&\checkmark &-\\
\midrule
&   Internal Producer &   \checkmark &     -&\checkmark &-&-&\checkmark            \\
\midrule
&   External Producer &   -&             
    -&\checkmark &-&-&\checkmark          
\\
\bottomrule
\end{tabular}
\begin{tablenotes}
    \footnotesize
    \noindent
    \begin{minipage}[c]{1\linewidth}
    \item [*] CUT: (implemented in) Class under test
    \item [**] GET method can be implemented in CUT or outside CUT
    \end{minipage} 
    \end{tablenotes}
\end{threeparttable}
\end{table*}

\subsubsection{Steps of the heuristic}\label{sec:heuristicExplanation}
Our heuristic contains three steps as explained below.

\paragraph{\textbf{Step 1}}
For each method call in a test case, step 1 of the heuristic is to collect information that should be assessed in order to verify the outcome(s) of the method call.
Overall, the information collected by step 1 depends on the method type (Section~\ref{sec: methClassification}) as explained further below, and summarized in Table~\ref{tab:informationCollection_Step1}.

\begin{itemize}
    \item A \textbf{creational} or \textbf{mutator} method is for initializing or modifying an object's attributes.
    The method might also return some value.
    Hence, to verify if a creational and mutator method behaves correctly, one needs to check if the attributes are modified correctly, and if the return value/object(s) (if any) is produced as expected. 
    Hence, the information to be collected is the modified attribute(s) and the return value/objects (if any). \\

    \item For a \textbf{get} (a sub type of \textbf{accessor}), we do not collect any information as we argue that it is a helper method in the context of unit testing, i.e., it is called to verify the outcome(s) of another method. \\

    \item An \textbf{internal producer} (a sub type of \textbf{accessor}) computes the return value/object based on an object's attribute(s) without modifying the object's attributes.
    Since the method's implementation (to produce the return value/object) is in the class under test, we argue that the purpose of invoking such a method is to verify if the implementation is correct.
    Hence, the information to be collected is the return value/object. \\

    \item An \textbf{external producer} (a sub type of \textbf{accessor}) is similar to an \textbf{internal producer} in the sense that the external producer also does not modify an object's state and the return value is computed based on the object's attribute(s).
    However,  the method's implementation (to produce the return value/object) is located outside of the class under test.
    Therefore, we argue that this method is invoked to check whether the object's attribute(s) is modified (or initialized) correctly by a mutator (or creational).
    In other words, we regard an \textbf{external producer} as a helper method in the context of unit testing, i.e., they are called to verify the outcome(s) of another method (similar to a \textbf{get} method).
    Hence, no information needs to be collected. \\
\end{itemize}

\begin{table}
\caption{Information to collect in Step 1 of the heuristic}
\label{tab:informationCollection_Step1}
\begin{tabular}{p{0.2\linewidth}p{0.15\linewidth}p{0.2\linewidth}p{0.22\linewidth}}
\toprule
Method Type & Retrieved attribute & Modified (or initialized)   attribute & Return value/object \\
\midrule
Creational        &   & \checkmark & {[}if any{]} \\
\midrule
Mutator           &   & \checkmark & {[}if any{]} \\
\midrule
Get               &   &   &                       \\
\midrule
Internal Producer &   &   & \checkmark            \\
\midrule
External Producer &   &   &                       \\
\bottomrule
\end{tabular}
\end{table}

\paragraph{\textbf{Step 2}}
Step 2 of the heuristic is to collect the information that \textbf{all} assert statements verify.
To acquire such information, we first collect the information that each assert statement verifies, then combine them together.

Similar to the logic behind Step 1, the information collected from each assert statement depends on the types of methods (Section~\ref{sec: methClassification}) invoked in the assert statement as explained further below, and summarized in Table~\ref{tab:informationCollection_Step2}).

\begin{itemize}
    \item If the assert statement invokes a \textbf{creational} method (or contains the object(s) created by a \textbf{creational} method call), what the assert statement tries to verify is whether the object is created correctly by the \textbf{creational}) method.
    Hence, the information to be collected is the created object and its attributes initialized by the \textbf{creational} method.
    
    \item If the assert statement invokes a \textbf{mutator} method (or contains return value/object(s) of a \textbf{mutator} method call), what the assert statement tries to verify is whether the attributes of the target object are modified correctly by the \textbf{mutator} method, and if the return value/object(s) (if any) is produced as expected.
    Hence, the information to be collected is the attributes modified by the \textbf{mutator} method AND the value/object(s) returned by the \textbf{mutator} method.    
    
    \item If the assert statement invokes a \textbf{get} method (or contains return values/objects of a \textbf{get} method call), we argue that developers do not try to check whether the \textbf{get} method is implemented correctly.
    Instead, the invocation of the \textbf{get} method is to verify the outcome(s) of another method.
    One typical example is that the assert statement assesses a get method's return value, which is an object's attribute, in order to verify if the attribute is modified correctly by a precedent mutator method.
    Hence, the information to be collected is the attribute(s) or object(s) returned by the \textbf{get} method.

    \item If the assert statement invokes an \textbf{external producer} method call (or contains return values/objects of an \textbf{external producer} method call), we argue that the purpose of invoking the \textbf{external producer} method is the same as invoking a get method.
    Hence, the information to be collected is the attribute(s) used by the \textbf{external producer} method to produce the return values or objects.
        
    \item If the assert statement invokes an \textbf{internal producer} method (or contains return values/objects of an \textbf{internal producer} method call), we argue that what the assert statement tries to verify is the behavior of the \textbf{internal producer} method only.
    Hence, the information to be collected is the return values/objects of the \textbf{internal producer} method.    
\end{itemize}

\begin{table}
\caption{Information to collect in Step 2 of the heuristic}
\label{tab:informationCollection_Step2}
\begin{tabular}{p{0.2\linewidth}p{0.15\linewidth}p{0.2\linewidth}p{0.22\linewidth}}
\toprule
Method Type & Retrieved attribute & Modified (or initialized)   attribute & Return value/object \\
\midrule
Creational        &   & \checkmark & {[}if any{]} \\
\midrule
Mutator           &   & \checkmark & {[}if any{]} \\
\midrule
Get               & \checkmark &   &              \\
\midrule
Internal Producer &   &   & \checkmark            \\
\midrule
External Producer & \checkmark &   &              \\
\bottomrule
\end{tabular}
\end{table}

At this step (Step 2), since we need to combine the information verified by every assert statement together, it is important to distinguish different changes to an attribute due to different method calls (creational, mutator methods).
In other words, the collected information should maintain the traceability between changes made to a modified (or initialized) attribute and the corresponding mutator (or creational) method calls.
To identify such a connection, we argue that if an attribute assessed by an assert statement is modified by multiple method calls, then among these method calls, the method call right before the assert statement is the one that the change of the attribute associates with. For example:

\begin{lstlisting}
public void test1() throws Exception{
    mutator1[attri1, attr2]
    mutator2[attr1]
    assertion1[attr1]
    mutator3[attr1, attr3]
    assertion2[attr1]
}
\end{lstlisting}
    
In this example, \textit{assertion1} assesses attribute \textit{attr1}, which is modified by mutator method \textit{mutator2}, and assertion2 assesses \textit{attr1}, which is modified by mutator method \textit{mutator3}.
To distinguish between the two connections: (1) \textit{mutator2} and \textit{attr1} \underline{assessed by \textit{assertion1}}, (2) \textit{mutator3} and \textit{attr1} \underline{assessed by \textit{assertion2}}, we update the attribute's name as follows before collecting the required information:

\begin{lstlisting}[escapechar=!] 
public void test1() throws Exception{
    mutator1[attr1, attr2]
    mutator2[attr1!\textcolor{red}{\textbf{a}}!]
    assertion1[attr1!\textcolor{red}{\textbf{a}}!]
    mutator3[attr1!\textcolor{brown}{\textbf{b}}!]
    assertion3[attr1!\textcolor{brown}{\textbf{b}}!]
}
\end{lstlisting}

\paragraph{\textbf{Step 3}}
Based on the information collected by Step 1 and Step 2, at Step 3 of the heuristic, we want to check if the information verified by all of the assert statements (Step 2) is the outcome(s) of one single method call (Step 1) in the test case.
If that is the case, then the test case is not an eager test; otherwise, the test case is classified as an eager test by the heuristic.

\subsubsection{Example}
In this section, we use a simple automatically generated test case extracted from Panichella et al.'s study~\cite{panichella2022test} as shown below to demonstrate how the heuristic works.

\begin{lstlisting}[escapechar=$]
@Test
  public void test2()  throws Throwable  {
      // $\color{DarkGreen}\(MethCall_1 - type: creational\)$  
      DirEntry dirEntry0 = new DirEntry();    

      // $\color{DarkGreen}\(MethCall_2 - type: mutator\)$
      dirEntry0.setSize((-1053L));  

      // $\color{DarkGreen}\(MethCall_3 - type: creational\)$
      DirEntry dirEntry1 = new DirEntry();  
      
      // $\color{DarkGreen}\(MethCall_4 - type: internal producer\)$
      boolean boolean0 = dirEntry0.equals((Object) dirEntry1);  

      // $\color{DarkGreen}\(Assert_1\) contains \(MethCall_5 - type: get\)$
      assertEquals((-1053L), dirEntry0.getSize()); 
      
      // $\color{DarkGreen}\(Assert_2\)$
      assertEquals(false, boolean0); 
  }
\end{lstlisting}

Based on the heuristic, for each method call in a test case, step 1 of the heuristic is to collect information that should be assessed in order to verify the outcome(s) of the method call.
Hence, the information collected in Step 1 is as follows:
\begin{align*}
    MethOutcome_1 &= \{dirEntry0\} \\
    MethOutcome_2 &= \{dirEntry0.size\} \\
    MethOutcome_3 &= \{dirEntry1\} \\
    MethOutcome_4 &= \{boolean0\} \\
    MethOutcome_5 &= \{dirEntry0.size\} \\
\end{align*}

Likewise, Step 2 of the heuristic is to collect the information that all assert statements verify.
To acquire such information, we first collect the information that each assert statement verifies, then combine them together.
Hence, the information collected in Step 2 is as follows:
\begin{align*}
    VerifiedInfo_1 &= \{dirEntry0.size\} \\
    VerifiedInfo_2 &= \{boolean0\} \\
    VerifiedInfo &= \{dirEntry0.size,\ boolean0\} \\    
\end{align*}

According to Step 3 of the heuristic, what needs to be checked is whether the information verified by all of the assert statements (Step 2) is the outcome(s) of one single method call (Step 1) in the test case.
Based on the information collected above, we can see that

\begin{align*}
    \forall i, VerifiedInfo &\not\subseteq\ MethOutcome_i  \\    
\end{align*}

This means that the test case \textbf{does not} verify the outcome(s) of a \textbf{single} method call; hence, it is an eager test.

\subsubsection{Limitations of the heuristic implementation}
This section presents the limitations when implementing our heuristic.

\paragraph{\textbf{Metaprogramming}}\mbox{} \\
Since our heuristic relies on static code analysis, it is not possible to work with metaprogramming, a programming technique in which a computer program has the ability to treat other programs as their data~\cite{lilis2019survey}.
In metaprogramming, we need a dynamic code analysis in order to identify the method types and collect the required data for the heuristic.
For example, the method below needs a dynamic analysis to know its type.
\begin{lstlisting}
	protected Object getBlobObject(String getterName) throws BeanBinException {
		try {
			Method getter = getGetter(getterName);			
			return getter.invoke(entity, new Object[0]);
		} catch (Exception e) {
			throw new BeanBinException("BlobClassTemplate.getBlobObject: " + e.getMessage(), e);
		}
	}
\end{lstlisting}

With the static code analysis, it is not possible to know what happens to the object ``entity" (the object of the class under test \textit{BlobClassTemplate}) when the method \textit{invoke} is called.
Consequently, the method type of \textit{invoke} cannot be identified.
It might still be feasible with human help, but it becomes difficult when automating the heuristic.

\paragraph{\textbf{External methods and objects outside class under test (CUT)}}\mbox{} \\
With external methods (methods outside of the production code), our heuristic assumes that they don't change the state of any object of CUT.
That is because it is not always possible to have access to the source code of such external methods.
Nonetheless, it is worth emphasizing that with our heuristic, it does not matter whether a method in the production code affects objects of CUT or outside of CUT.
As long as the affected object can be identified, it should be possible to collect the related information required by the heuristic.

\subsubsection{Scalability of the heuristic}
The heuristic relies on the static analysis of the source code, which might slow down its performance.
Not being able to deal with metaprogramming also restricts its scalability.
While the current scope of the heuristic is limited to Java, the heuristic can be extended to other object-oriented programming languages.
In our recent work, we are also developing a tool that operationalizes the proposed heuristic to further boost scalability by reducing the manual effort.

\subsection{Pseudocode for the heuristic}
\label{sec:heuristicPseudocode}
In this section, we present our heuristic in pseudocode as shown in Pseudocode~\ref{alg:heuristic_pseudocode}.

\begin{algorithm*}
\caption{Our heuristic to detect eager test}\label{alg:heuristic_pseudocode}
\begin{algorithmic}[1]
\Let $MethCalls$ be a list containing all method calls in the test case
\Let $Asserts$ be a list containing all assertion statements in the test case
\Statex
\State \textbf{begin} \textit{Step 1}:
\State Initialize array $MethOutcomes \gets \emptyset$
\Foreach {$\text{method call} ~MethCall_i \in MethCalls$}
    \State Identify the method type of $MethCall$\(\sb{i}\) based on the descriptions of method types
    \State Initialize $MethOutcome_i \gets \emptyset$
    \Statex
    \If{$\text{type of} MethCall_i ~\text{is \textit{Get} OR \textit{External Producer}}$}
        \State do nothing
    \ElsIf{$\text{type of} MethCall_i ~\text{is \textit{Creational}}$}
        \State $MethOutcome_i \gets \text{all attributes initialized by} MethCall_i$
        \If{$MethCall_i$ returns any value or object}
            \State $MethOutcome_i \gets MethOutcome_i \cup ~\text{all return values and objects of} ~MethCall_i$
        \EndIf
    \ElsIf{$\text{type of} MethCall_i ~\text{is \textit{Mutator}}$}
        \State $MethOutcome_i \gets \text{all attributes modified by} ~MethCall_i$
        \If{$MethCall_i$ returns any value or object}
            \State $MethOutcome_i \gets MethOutcome_i \cup ~\text{all return values and objects of} ~MethCall_i$
        \EndIf
    \ElsIf{$\text{type of} MethCall_i ~\text{is \textit{Internal Producer}}$}
        \State $MethOutcome_i \gets \text{all return values and objects of} ~MethCall_i$
    \EndIf
\EndFor
\State Add $MethOutcome_i$ to $MethOutcomes$
\State \textbf{end} \textit{Step 1}
\Statex

\State \textbf{begin} \textit{Step 2}:
\State Initialize $VerifiedInfo \gets \emptyset$
\Foreach{$\text{assert statement} ~Assert_j \in Asserts$}
    \State $VerifiedInfo_j \gets \text{all objects and attributes passed as arguments in} ~Assert_j$
    \State $MethCalls\_Assert_j \gets ~\text{all method calls in} ~Assert_j$
    \Statex
    \Foreach {$\text{method call} ~MethCall_k \in MethCalls\_Assert_j$}
        \State Identify the method type of $MethCall_k$ based on the descriptions of method types        
        \Statex
        \If{$\text{type of} MethCall_k ~\text{is \textit{Get} OR \textit{External Producer}}$}
            \State $VerifiedInfo_j \gets VerifiedInfo_j \cup \text{all attributes retrieved by} MethCall_k$
        \ElsIf{$\text{type of} MethCall_k ~\text{is \textit{Creational}}$}
            \State $VerifiedInfo_j \gets VerifiedInfo_j \cup \text{all attributes initialized by} MethCall_k$
            \If{$MethCall_k$ returns any value or object}
                \State $VerifiedInfo_j \gets VerifiedInfo_j \cup ~\text{all return values and objects of} ~MethCall_k$
            \EndIf
        \ElsIf{$\text{type of} MethCall_k ~\text{is \textit{Mutator}}$}
            \State $VerifiedInfo_j \gets VerifiedInfo_j \cup \text{all attributes modified by} ~MethCall_k$
            \If{$MethCall_k$ returns any value or object}
                \State $VerifiedInfo_j \gets VerifiedInfo_j \cup ~\text{all return values and objects of} ~MethCall_k$
            \EndIf
        \ElsIf{$\text{type of} MethCall_i ~\text{is \textit{Internal Producer}}$}
            \State $VerifiedInfo_j \gets VerifiedInfo_j \cup \text{all return values and objects of} ~MethCall_k$
        \EndIf
    \EndFor
    \State $VerifiedInfo \gets VerifiedInfo \cup VerifiedInfo_j$
\EndFor
\State \textbf{end} \textit{Step 2}
\Statex
\algstore{myalg}
\end{algorithmic}
\end{algorithm*}

\begin{algorithm*}                     
\begin{algorithmic} [1]
\algrestore{myalg}
\State \textbf{begin} \textit{Step 3:}
\State Initialize $isEagerTest \gets false$
\State Initialize $NumOfVerifiedMeths \gets 0$
\ForAll{$MethOutcome_l$ in $MethOutcomes$}
    \If {$VerifiedInfo \not\subseteq\ MethOutcome_l$}
        \State $isEagerTest \gets true$
        \State \textbf{Return} $isEagerTest$
    \EndIf
\EndFor
\For{$l \gets 1\, \textbf{to}\, m$}
    \If{$VerifiedInfo \subseteq MethOutcome_l$}
        \State $NumOfVerifiedMeths \gets NumOfVerifiedMeths + 1$
    \EndIf

    \If{$NumOfVerifiedMeths \geq 2$}
        \State $isEagerTest \gets true$
        \State \textbf{Return} $isEagerTest$
    \EndIf
\EndFor
\State \textbf{return} $isEagerTest$
\State \textbf{end} \textit{Step 3}
\end{algorithmic}
\end{algorithm*}
\pagebreak

\section{Comparison with existing rules (RQ5)}\label{sec:resultAnalysis_RQ5}
In this section, we first present the detection rules selected for comparison. Then, we present our findings regarding the levels of agreement between each pair of detection rules, followed by our findings regarding the agreement levels between the heuristic and each detection rule.

\subsection{Existing detection rules selection outcome}\label{sec:selectedDetectionRules}
From the literature review, we identified in total six rules to detect eager tests.
Details of the rules are in Section~\ref{sec:resultAnalysis_RQ2} and Table~\ref{tab:eagerTest_detectionRules}.
Among these six rules, we could not apply two of them in our manual analysis.
The first one, proposed by Bavota et al.~\cite{Bavota2012empirical}, works at the test suite level, while we wanted to detect eager tests at the test case level instead.
The second rule, proposed by Palomba et al.~\cite{Palomba2018Automatic}, relies on the textual similarity among the tested methods in a test case to classify if the test case is an eager test.
The rule was excluded since calculating such similarity could not be done manually.

Consequently, we had four detection rules from the literature to compare with our heuristic.
The selected rules were implemented in four~\cite{breugelmans2008testq, Bleser2019SoCRATES, Peruma2020TsDetect, virginio2020jnose} out of the seven existing tools~\cite{Aljedaani2021Test}, capable of detecting eager tests, while the excluded rules formed the basis for the remaining three tools~\cite{Lambiase2020Just, Palomba2018Automatic, Bavota2012empirical}.
Thus, we believe our study has comprehensively considered all relevant detection rules and tools available in the literature.

Table~\ref{tab:comparison_detectionRules} presents the selected rules for comparison with our heuristic.
Note that one of these four rules comes with three different thresholds.
Hence, we decided to consider each threshold as an independent rule to achieve a thorough comparison.
It is also worth emphasizing that we studied the primary studies proposing these rules as well as the source code of the detection tools that implemented the rules to make sure that we understood the rules correctly. 
The information in column \textbf{Notes} of Table~\ref{tab:comparison_detectionRules} summarizes the extra details we learned from this extra step.

\begin{table*}
{    
    \footnotesize
    \begin{center}
    \caption{Eager test detection rules from the literature to compare with our heuristic}
    \label{tab:comparison_detectionRules}       
    \begin{threeparttable}    
    \begin{tabular}{p{0.04\textwidth}p{0.4\textwidth}p{0.4\textwidth}p{0.03\textwidth}}
        \toprule
        \textbf{ID} & \textbf{Detection rule} & \textbf{Notes} & \textbf{Ref} \\
        \midrule
        DR1 & At least 2 method calls from \textbf{CUT} in a test case & Excluding constructor & \cite{Bleser2019Assessing} \\
        DR2.1 & At least 3 PTMI in a test case & Excluding constructor, considering only methods from \textbf{CUT} & \cite{breugelmans2008testq} \\
        DR2.2 & At least 5 PTMI in a test case & Excluding constructor, considering only methods from \textbf{CUT} & \cite{Peruma2020TsDetect} \\
        DR2.3 & At least 5 PTMI in a test case & Excluding constructor, considering only methods from \textbf{production code} \textit{(code developed by the project team that will end up in the released product)} & \cite{Rompaey2007On} \\           
        DR3 & At least two cycles of non-verification instructions followed by verification instructions & None & \cite{Pizzini2022Automatic} \\
        DR4 & At least 2 assertions in a test case and at least one assertion is not on the result of a get method & No definition of get methods & \cite{Panichella2020Revisiting} \\
        \bottomrule
    \end{tabular}
    \begin{tablenotes}
        \footnotesize
        \noindent
        \begin{minipage}[c]{1\linewidth}
            \item CUT: Class under test
            \item PTMI: production type method invocations
        \end{minipage} 
    \end{tablenotes}
    \end{threeparttable}
    \end{center}
}
\end{table*}

\begin{figure*}
\begin{center}
\includegraphics[width=0.9\linewidth]{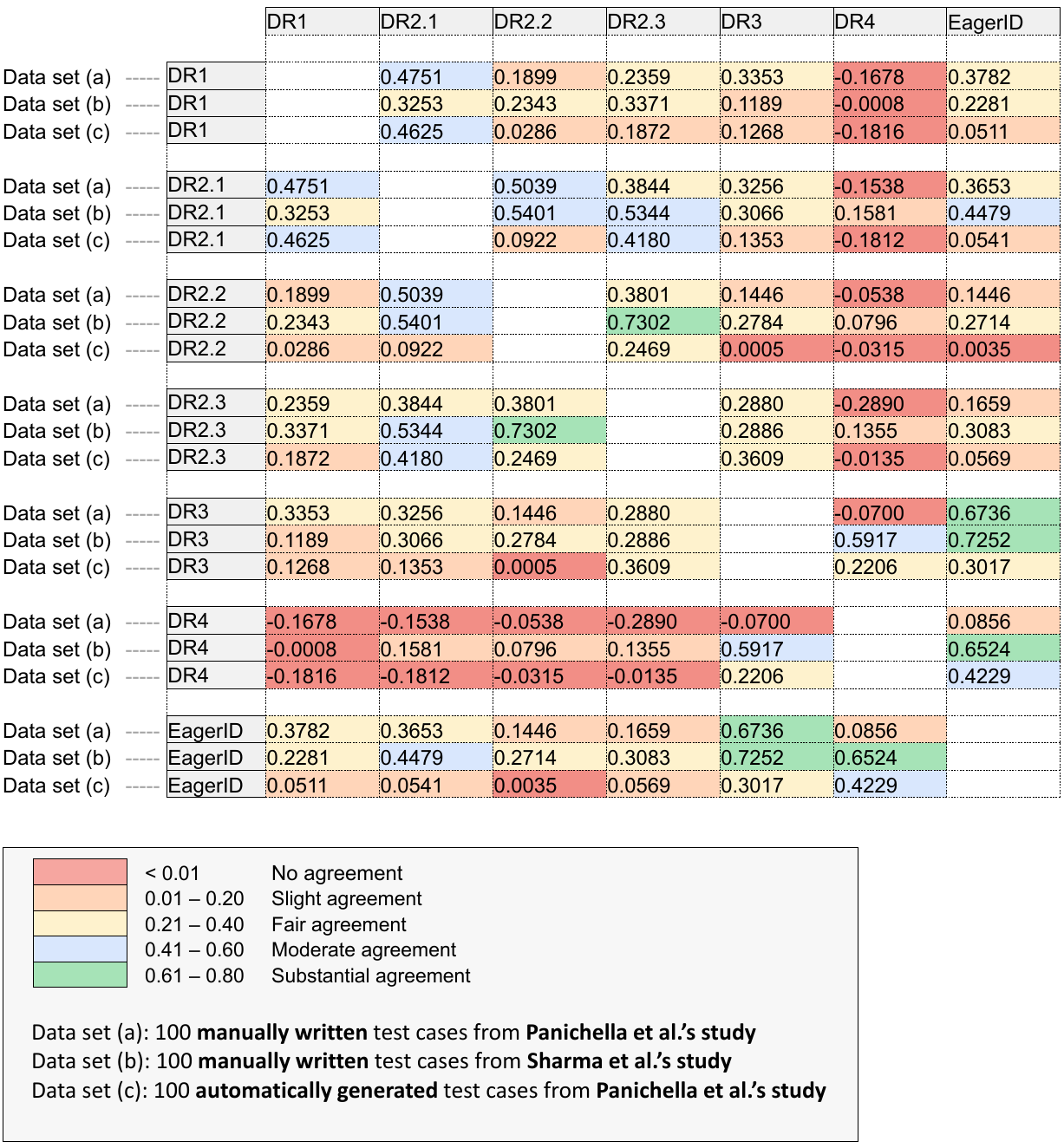}
\caption{Degree of agreement (using Cohen's kappa coefficient) in the eager test classification between the detection rules in the literature (Table~\ref{tab:comparison_detectionRules}) and the proposed heuristic (EagerID)}
\label{fig:cohen_kappa}
\end{center}
\end{figure*}

Figure~\ref{fig:cohen_kappa} presents Cohen's kappa calculated for every pair of the detection rules (DR1 -- DR4) and between our heuristic (EagerID) and each detection rule.
Note that we call our heuristic \textit{EagerID} in Figure~\ref{fig:cohen_kappa} for the sake of simplification.

Our manual assessment involved three separated data sets: (a) 100 manually written test cases from Panichella et al.'s study, (b) 100 manually written test cases from Sharma et al.'s study, and (c) 100 automatically generated test cases from Panichella et al.'s study (details in Section~\ref{sec:testCaseSelection}).
Hence, the left side of Figure~\ref{fig:cohen_kappa} indicates which data set the kappa values were calculated from.
For example, for the data set (a), the first row of Figure~\ref{fig:cohen_kappa} shows the kappa values between two detection rules, DR1  and DR2.1, as 0.4751.

Also note that in Figure~\ref{fig:cohen_kappa}, we presented the kappa values according to each detection rule and the heuristic, instead of showing them in three tables, one for each data set.
The purpose is to provide a better overview of how each detection rule and the heuristic agree with each other in terms of identifying eager tests from all data sets.
For example, the first three rows in Figure~\ref{fig:cohen_kappa} present the kappa values between the detection rule DR1 and each of the following: DR2.1, DR2.2, DR2.3, DR4, and the heuristic, from each data set.
Based on Landis and Kock's guideline~\cite{landis1977measurement}, we observed five levels of agreement ranging from ``no agreement'' (in red) to ``substantial agreement'' (in green) as shown in Figure~\ref{fig:cohen_kappa}.

\subsection{Agreement between the selected detection rules}
Overall, Cohen's kappa for each pair of the selected detection rules varied from ``no agreement'' to ``substantial agreement'' (Figure~\ref{fig:cohen_kappa}).
It indicates a significant misalignment between the selected rules in detecting eager test cases.
Nevertheless, looking more closely at the kappa values, we could see that some detection rules tended to have higher levels of agreement with certain detection rules than with the other rules.

With the detection rule DR1, its agreement with the detection rule DR2.1 was much higher than with the other rules.
It is reasonable as both DR1 and DR2.1 share a common principle, i.e., counting the number of method calls from the class under test, and their thresholds are not so different from each other (2 method calls for DR1 and 3 method calls for DR2.1).
Meanwhile, although DR2.2 and DR2.3 are also based on the number of method calls, their thresholds (5 method calls) were much higher than DR1 and DR2.1.
Therefore, it makes sense that DR1 has less agreement with DR2.2 and DR2.3.

One interesting observation is that even though the detection rule DR3 is not based on the same principle as DR1, the levels of agreement between DR1 and DR3 for all data sets were not distinctively different from the agreement levels between DR1 and DR2.1 or between DR1 and DR2.2. 
One possible reason is that DR3 requires at least two cycles of non-verification instructions followed by verification instructions in a test case.
A cycle of non-verification instructions is likely to involve at least one method call.
Hence, an eager test based on DR3 is more likely to have at least two method calls, which fulfills the requirement of DR1.

While DR1 shared some agreement (even though not very high) with DR2.1, DR2.2, DR2.3, and DR3, as analyzed above, DR1 had no agreement with DR4 for all data sets.
It is probably due to the fact that DR1 and DR4 are based on completely different principles. (DR4 requires at least two assertions, and at least one assertion should not be on the result of a get method.)
Hence, their eager test detection outcomes were not aligned with each other.

Next, we looked at the agreement levels between the three detection rules DR2.1, DR2.2, and DR2.3.
Since these rules share the same principle, that is, counting the number of method calls, it is reasonable that their agreement was relatively high compared to the rest.
For example, while both DR2.1 and DR2.2 count the number of method calls from the class under test, the threshold of DR2.1 was three method calls and the threshold of DR2.2 was five method calls.
Hence, if DR2.2 classified a test case as an eager test, DR2.1 would naturally give the same assessment.
With the detection rule DR2.3, as it requires at least five method calls from the production code, DR2.3 covers DR2.2 which requires at least five method calls from a smaller scope, i.e., the class under test.
It means that test cases classified as eager tests by DR2.2 would be covered by DR2.3 as well.
We argue that their only differences in threshold and the scope of the counted methods caused the misalignment in their detection outcomes.
Meanwhile, this group of detection rules had either no agreement or, at most, slight agreement with the detection rule DR4.
It is probably because DR4 is based on a completely different principle than DR2.1, DR2.2, and DR2.3 as explained earlier.

While the detection rule DR4 appeared to have high levels of disagreement with the rest, it had the least conflict with DR3.
We believe that is because of an implicit common factor in these two rules.
More specifically, DR3 requires at least two cycles of non-verification instructions followed by verification instructions in a test case, which implies at least two assertions.
Having at least two assertions is also a requirement of DR4.
However, DR4 put more restrictions on the assertions by requiring that at least one assertion must be not on the result of a get method.
It led to the disagreement between DR4 and DR3.

\subsection{Agreement between our heuristic and the selected detection rules}
In the previous section, we presented our findings regarding the agreement levels between each pair of the detection rules.
Here, we look at the extent to which our heuristic agreed with each detection rule in terms of identifying eager tests from the same data sets: (a) 100 manually written test cases from Panichella et al.'s study, (b) 100 manually written test cases from Sharma et al.'s study, and (c) 100 automatically generated test cases from Panichella et al.'s study (details in Section~\ref{sec:testCaseSelection}).

\begin{figure*}
\begin{center}
\includegraphics[width=0.9\linewidth]{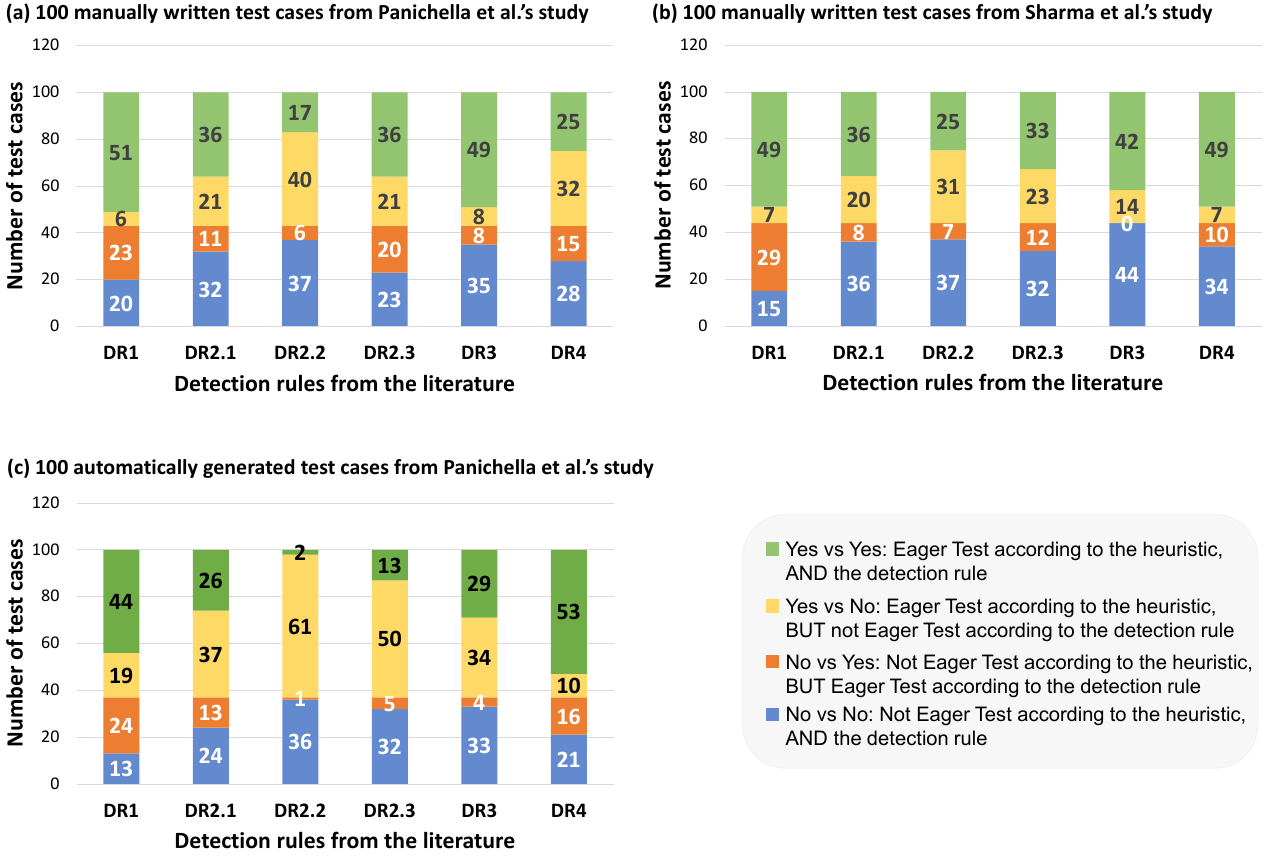}
\caption{Eager test classification based on our heuristic and the detection rules in the literature (Table~\ref{tab:comparison_detectionRules})}
\label{fig:comparison_heuristic_detectionRules}
\end{center}
\end{figure*}

Figure~\ref{fig:comparison_heuristic_detectionRules} shows the percentages of test cases classified as eager tests or not based on our heuristic and each detection rule.
For example, on the one hand, 23\% of the manually written test cases from Panichella et al.'s study were classified as eager tests according to the detection rule DR1 (At least 2 method calls from class under test (CUT) in a test case) but not by our heuristic.
On the other hand, our heuristic marked 6\% of the same data set as eager tests, while the detection rule DR1 did not.
To better understand the agreement levels between our heuristic and each detection rule, we present the calculated kappa between our heuristic and each detection rule in Figure~\ref{fig:cohen_kappa}.
Overall, based on both Figure~\ref{fig:cohen_kappa} and Figure~\ref{fig:comparison_heuristic_detectionRules}, we observed the same pattern as between the detection rules, that is, there was little agreement between our heuristic and each rule.

First, between the two types of test cases (manually written versus automatically generated), the eager test classification for the automatically generated test cases had more disagreement.
One factor that might potentially explain this observation is that the automatically generated test cases often follow one pattern as they were all created by EVOSUITE.
For example, it appeared very frequently that EVOSUITE inserted an assert statement to verify the outcome of a creational method call in the Set-up of a test case.
Naturally, the test case also has at least another assert statement in the Verify part to verify the outcome(s) of another method call.
The test case, therefore, tries to verify the outcomes of multiple method calls, which is classified as an eager test, according to our heuristic.
However, the other detection rules from the literature do not have the same conclusion.
On the contrary, the manually written test cases were created by different developers with different styles or mindsets when designing test cases.
Hence, their test cases are more diverse than the automatically generated ones.

Second, we also noticed that there are some detection rules having higher levels of agreement with our heuristic than others.
For all data sets, there was either no agreement or, at most, fair agreement between the detection rule DR2.2 (At least 5 method calls from CUT in a test case) and our heuristic.
It makes DR2.2 have the highest disagreement with our heuristic.
In contrast, the detection rule DR3 (At least two cycles of non-verification instructions followed by verification instructions) was the one having the least disagreement with our heuristic, especially in the case of the manually written test cases (with substantial agreement).
Meanwhile, the levels of agreement between our heuristic and two detection rules DR2.1 and DR4 fluctuated, ranging widely from slight agreement to moderate agreement (with DR2.1), or from slight agreement to substantial agreement (with DR4). We further discuss the root causes of the differences between our heuristic and the selected detection rules in Section~\ref{sec:discussion} and their implications in Section~\ref{sec:discussion_implications}.

\subsection{Agreement between our assessment and the assessments from the studies of Panichella et al.~\cite{panichella2022test} and Sharma et al.~\cite{tushar2023investigating}}\label{sec:comparisonWithPanichellaAndSharma}
With Panichella et al.'s study, as explained earlier in Section~\ref{sec:testCaseSelection}, the authors proposed the detection rule DR4.
However, since they analyzed the eagerness of test suites, not test cases, we could not conduct a direct comparison in the assessment results.
Nevertheless, the authors did mention that 39 out of the 49 manually written test suites (approximately 80\%) that they assessed were eager according to the detection rule DR4.
Meanwhile, our manual assessment showed that 57\% of the assessed test cases (manually written) were eager tests.
In terms of automatically generated test cases, also based on DR4, Panichella and his colleagues found that 20 out of the 100 test suites were eager.
On the contrary, our heuristic classified 63 out of 100 test cases as eager tests.
These differences show that the eagerness of test suites alone does not reflect real, rectifiable issues in test cases.
On top of that, since our assessment also demonstrates the misclassification problem of the detection rule DR4, their assessment result might not be a reliable indicator for eager tests in practice.

In Sharma et al.'s study, the authors used the detection rule DR2.3 in their manual assessment.
Our assessment agreed with Sharma et al. in terms of the test cases that they marked as eager tests, which are 13 in total.
However, there were 33 test cases (out of 100) that were classified as eager by our heuristic but discarded by Sharma and his colleagues.
This disagreement probably stems from the fact that we had no clear description regarding whether they actually considered any method call from the production code (as we assumed) or strictly limited to the class under test.
Our second assumption is that their threshold was five method calls based on the test cases they marked as eager.
This second assumption was made due to no extra details provided in their paper.

\section{Discussion}\label{sec:discussion}
In this section, we discuss the main aspect of our results, that is the root causes of the differences between our heuristic and the selected detection rules.

To understand the reasons behind the misalignment in detecting eager tests between our heuristic and each detection rule, we looked into the test cases that drew different assessment outcomes.
Our investigation was in a pair-wise basic, i.e., between the heuristic and each of the detection rules.

According to the detection rule DR1, as long as a test case has only one method call from the class under test (CUT), it is not an eager test.
However, the assessed test cases showed that even if a test case meets the condition, it can still be an eager test when having one or more assert statements in the Set-up (besides the assert statements for verifying the method call from CUT).
Generally, the Set-Up contains the instantiation of the object under test which might require the invocations of methods (also known as helper methods) to bring the object under test to the desired state for testing.
Hence, we argue that when a test case has some extra assert statements to verify the outcomes of method calls in the Set-up, the test case no longer verifies the outcome(s) of one single method call, which consequently should be classified as an eager test instead.
One example test case to demonstrate this scenario is as follows:
\begin{lstlisting}
    Example Test Case #1
    @Test
    public void test3()  throws Throwable  {
        Home home0 = new Home();
        PhotoController photoController0 = new PhotoController(home0, (UserPreferences) null, (View) null, (ViewFactory) null, (ContentManager) null);
        assertNotNull(photoController0);
      
        long long0 = photoController0.getTime();
        assertEquals(1372766400496L, long0);
    }
\end{lstlisting}

Another reason that makes this detection rule problematic is that it does not count constructors in the assessment.
In other words, the rule does classify a test case that contains multiple assert statements verifying multiple constructors as an eager test.
Nevertheless, we argue that this scenario satisfies the original definitions of eager tests and should not be discarded.
One example is as follows:
\begin{lstlisting}
    Example Test Case #2
    @Test
    public void testConstr() throws Exception {
        HttpBot bot = new HttpBot(getValue("wikiMW1_13_url"));
        assertNotNull(bot);
        bot = new HttpBot(new URL(getValue("wikiMW1_13_url")));
        assertNotNull(bot);
    }
\end{lstlisting}

Our assessment also showed that the detection rule DR1 could wrongly classify a non-eager test, i.e., a test case is not an eager test even though it contains at least two method calls from CUT.
This falls into two typical scenarios.
The first one is when a test case has multiple method calls from CUT in its Set-up and/or Teardown while having one single assert statement verifying a single method call from CUT.
These method calls in the Set-up and/or Teardown are the helper methods that should not be considered a sign of eagerness of the test case.
The following test case can demonstrate our point:
\begin{lstlisting}
    Example Test Case #3
    @Test
    public void testGetConnectionUserPassSetters() throws Exception {
        String username = "user";
        String password = "_secret";
        String url = "jdbc:h2:mem:ormlite-up;USER=" + username + ";PASSWORD=" + password;
        JdbcConnectionSource sds = new JdbcConnectionSource(url);
        sds.setUsername(username);
        sds.setPassword(password);
        assertNotNull(sds.getReadOnlyConnection(null));
        sds.close();
    }
\end{lstlisting}

The second scenario is when a test case contains multiple \textbf{get} methods (defined and discussed in Section~\ref{sec: methClassification}).
These \textbf{get} methods are from CUT but invoked to verify the outcome(s) of a single method call from CUT.
Clearly, such invocations of \textbf{get} methods, which are very common in practice, do not make the test case an eager test.
The test case below falls into this scenario:
\begin{lstlisting}
    Example Test Case #4
    @Test
    public void testAccessors() throws Exception {
        final String uri = "http://localhost:80";
        final String proxyHost = "foo";
        final int proxyPort = 1030;
        final String proxyUser = "bar";
        final String proxyPassword = "fly";
        final SSLProperties ssl = new SSLProperties();

        HTTPRequestInfo info = populate(uri, proxyHost, proxyPort, proxyUser,
                                        proxyPassword, ssl);

        assertEquals(uri, info.getURI().toString());
        assertEquals(proxyHost, info.getProxyHost());
        assertEquals(proxyPort, info.getProxyPort());
        assertEquals(proxyUser, info.getProxyUser());
        assertEquals(proxyPassword, info.getProxyPassword());
        assertEquals(ssl, info.getSSLProperties());
    }    
\end{lstlisting}

The detection rules DR2.1 and DR2.2 are not different from DR1 except for the threshold.
While DR2.1 states that a test case must have at least three method calls from CUT to be classified as an eager test, DR2.2 increases the threshold to five method calls from CUT.
Since the thresholds of DR2.1 and DR2.2 are much higher than DR1, the number of test cases classified as eager tests by DR2.1 and DR2.2 is much lower than by DR1 (as shown in Figure~\ref{fig:comparison_heuristic_detectionRules}).
In contrast, DR2.1 and DR2.2 missed more eager tests than DR1 due to the same reason.
For example, the following test case is classified as an eager test by DR1 (as well as by our heuristic) but not by DR2.1 and DR2.2 due to the differences in threshold.
\begin{lstlisting}
    Example Test Case #5
    @Test
    public final void testEntities() {
        String s = "&#039;";
        String t = "'";
        assertEquals(t, MediaWiki.decode(s));
        s = "&quot;";
        t = "\"";
        assertEquals(t, MediaWiki.decode(s));
    }
\end{lstlisting}

In the same group of detection rules, DR2.3 is less restrictive than DR2.1 and DR2.2, as DR2.3 considers any method call from the production code while having a threshold of 5.
However, our manual assessment showed that DR2.3 still misclassified many test cases.
As the concept behind DR2.3 is actually the same as DR1, DR2.1, and DR2.2, the misclassified test cases by DR2.3 also fall into the two scenarios that we explained earlier with DR1.

The detection rule DR3 classifies eager tests from a different angle.
According to this rule, an eager test must have at least two cycles of non-verification instructions followed by verification instructions.
The first problem of this detection rule is that even if a test case has only one cycle of non-verification instructions followed by verification instructions, it can still be an eager test when containing multiple assert statements verifying the outcomes of multiple method calls.
For example, the test case below has only one cycle of non-verification instructions followed by verification instructions.
However, it verifies multiple method calls; hence, it should not be discarded.
\begin{lstlisting}
    Example Test Case #6
    @Test
    public void testEquals() throws Exception {
        Category a = new Category();
        a.setName("giraffes");

        Category b = new Category();
        b.setName("camels");

        Category c = new Category("giraffes");

        assertEquals(a, c);
        assertEquals(c, a);
        assertTrue(!a.equals(b));
        assertTrue(!b.equals(c));
        assertTrue(!b.equals(null));
        assertTrue(!b.equals(new Object()));
    }
\end{lstlisting}

The second problem of this detection rule is that a test case with at least two required cycles might not be an eager test.
For example, a test case can contain an assert statement verifying the outcome(s) of one method call from either CUT or production code in the first cycle.
In the second cycle, one or more \textbf{get} methods are called to acquire extra information regarding the outcome(s) of the method call in the first cycle.
This test case is not an eager test as all the assert statements verify the outcome(s) of one single method call.
For example, one test case falls into this scenario is shown below.
\begin{lstlisting}
    Example Test Case #7
    @Test
    public void testBasic() throws Exception {
        BeanBinDAO dao = getDAO();
                
        TestEntity entity = new TestEntity();
        entity.setAnInt(5);
        entity.setString("a sample string");
                
        Transaction tx = new Transaction(dao, TestEntity.class);
        tx.addCommand(new AddEntity(entity));
        tx.commit();
                
        Query q = new Query(new Criteria("string", "a sample string", SearchType.EQUALS));
        List<Object> results = dao.search(TestEntity.class, q);
                
        assertEquals(1, results.size());
        TestEntity saved = (TestEntity) results.get(0);
        assertEquals("a sample string", saved.getString());
    }
\end{lstlisting}

The last detection rule in our manual assessment was DR4, which emphasizes that an eager test needs to have at least two assert statements, and at least one of the assert statements is not based on a \textbf{get} method's result.
The first thing to note is that this detection rule does not come with a clear description of \textbf{get} methods.
A get method can be a simple getter such as \textit{getName()}, or it can be something less traditional such as \textit{isEmpty()}.
In our assessment, we used our definition of \textbf{get} methods for this detection rule, i.e., a get method returns an object's attribute without modifying the attribute.
Like the other detection rules discussed above, DR4 can also miss detecting eager tests.
It occurs when a test case contains multiple assert statements, and each of the assert statements invokes a get method to verify the outcome(s) of a different method like a mutator or a constructor, etc.
The test case clearly satisfies the original definitions of the Eager Test smell while being missed by DR4.
One example is as follows:
\begin{lstlisting}
    Example Test Case #8
    @Test
    public void testGetSetter() throws Exception {
        Method getter = EntityUtils.getMethod(IndexedEntity.class, "keywords");
        Method setter = EntityUtils.getSetter(getter);
        assertEquals("setKeywords", setter.getName());
        
        getter = EntityUtils.getMethod(IndexedEntity.class, "generatedKeywords");
        setter = EntityUtils.getSetter(getter);
        assertEquals(null, setter);
        }
\end{lstlisting}

This detection rule DR4 can also falsely flag a non-eager test.
The common scenario we learned from our manual assessment is that a test case has at least two assert statements, and at least one invokes an \textit{external producer} instead of a \textbf{get} method.
Such external producers can be \textit{toString()} or \textit{contains()}.
These external producers act like helper methods, i.e., to extract information regarding the outcome(s) of a single method call.
Hence, in this scenario, the test case is actually not an eager test, as shown in the example below.
\begin{lstlisting}
    Example Test Case #9
    @Test
    public void testFindImplementations() throws Exception {
        Resolver resolve = new Resolver();
        List<Class> list = resolve.findImplementations(Base.class);
        assertTrue(list.contains(ImplOne.class));
        assertTrue(list.contains(ImplTwo.class));
        assertEquals(2, list.size());
        }
\end{lstlisting}

\subsection{Implications}\label{sec:discussion_implications}



Through our manual analysis, we have identified multiple patterns of test cases that often appear in practice but are misclassified by the detection rules.
Basically, there are three common patterns where non-eager test cases are misclassified as eager tests by the widely-used detection rules. These include test cases that (1) require multiple method calls for setup but verify only one method, (2) use multiple \textbf{get} methods to validate the outcome(s) of a single method call, or (3) invoke helper methods that are not part of the production code to acquire information regarding the outcome(s) of a single method call such as \textit{size()}, \textit{contains()}.
Additionally, our analysis highlights two patterns of eager tests that are missed by these rules but detected by the proposed heuristic: (1) tests verifying multiple constructors and (2) tests with multiple get methods validating the outcomes of different methods.

Given the high occurrence of these patterns in practice, their misclassification significantly affects detection results.
This raises concerns about the reliability of current tools, potentially undermining both practitioner trust and academic claims regarding the prevalence and impact of eager tests.


Ultimately, the main reason for the misclassification of these common patterns of test cases is that the existing rules do not take into account what a method actually does, which is the determining factor for eager test recognition.
The question here is whether the authors of these rules and tools took this limitation into account while proposing the rules and implementing the detection tools.
In fact, we have not seen any studies where any limitations of these rules and tools were well documented.
Only Bavota et al.~\cite{Bavota2015Are}, whose proposed detection rule is ``JUnit classes having at least one method that uses more than one method of the tested class'', explained briefly that their rule is ``very simple that overestimates the presence of test smells in the code''.
Their reason was to not miss any test smell instances.
Our heuristic, in contrast, analyzes the purpose of each method call by systematically studying the production code and the test code based on the method type before drawing conclusions.
Therefore, even though it might require more effort to implement our heuristic due to the very step of analyzing the code, we believe that the trade-off is worth it as it is important to reduce such a high misclassification rate in terms of eager tests.

\section{Conclusion and future work}\label{sec:conclusion}
Eager Test has been one of the most common test smells discussed by many researchers.
There are over ten test smell detection tools, including the Eager Test smell in their test smell lists.
Nevertheless, the existing detection rules and approaches used by those detection tools are all based on inadequate interpretations of the Eager Test smell.
Due to this reason, conclusions on the prevalence and the negative impacts of the Eager Test smell in practice might become invalid.
To address this problem, we conducted a literature review to obtain a better understanding of the definitions of the Eager Test smell as well as the existing detection rules and approaches for this test smell.
Based on the information obtained from the literature review, we proposed a novel, unambiguous definition of the Eager Test smell and, accordingly, a heuristic (with pseudocode) to detect eager tests in unit test cases written in Java.
We also performed a manual assessment to analyze how the heuristic could overcome the shortcomings of the existing detection rules.

Overall, the contribution of this study is two-fold: (1) the original definitions of the Eager Test smell together with the current detection rules found in the literature, and (2) the novel, unambiguous definition of the Eager Test smell and, accordingly, a heuristic (with pseudocode) to detect eager tests.
Our findings provide researchers with an overview of how the Eager Test smell has been defined and detected in the literature.
On top of that, the manual assessment of the proposed heuristic highlights the importance of having a more precise approach to detect eager tests in the context of unit testing and the concrete shortcomings of the existing detection rules.
In our recent work, we are also developing a tool, namely EagerID \footnote{\url{https://github.com/vi-tran1987/ET_heuristic}}, that operationalizes the proposed heuristic.
For future work, we plan to use our tool to conduct automatic assessment of eager tests on a larger scale, including test cases from industrial projects, to further validate our heuristic in real-world scenarios.
Additionally, we aim to compare our heuristic with other prominent test smell detection tools for Java.
We also plan to explore defining the Eager Test smell for non-object-oriented programming languages, such as C, to broaden the applicability of our approach across diverse programming paradigms.

\section*{Acknowledgments}
This work has been supported by ELLIIT; the Strategic Research Area within IT and Mobile Communications, funded by the Swedish Government. The work has also been supported by a research grant for the GIST (reference number 20220235) and SERT project from the Knowledge Foundation in Sweden.

\begin{appendices}
\section{Comparison of the outcomes for Eager test classifications of our heuristic (EagerID) and existing detection rules}
\label{appendix:ET_classification_outcome}
\newpage\hbox{}\thispagestyle{empty}
\begin{figure*}[htb]
\begin{center}
\includegraphics[width=1\linewidth]{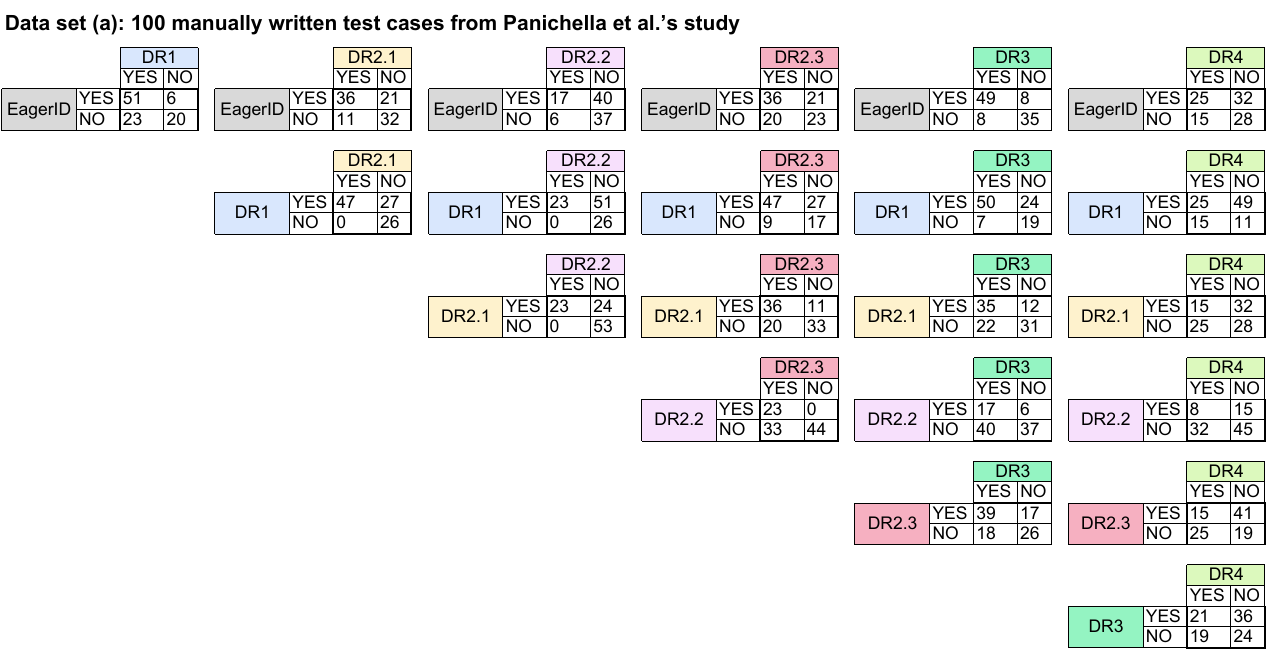}
\label{fig:ET_outcome_dataSetA}
\end{center}
\end{figure*}

\begin{figure*}[htb]
\begin{center}
\includegraphics[width=1\linewidth]{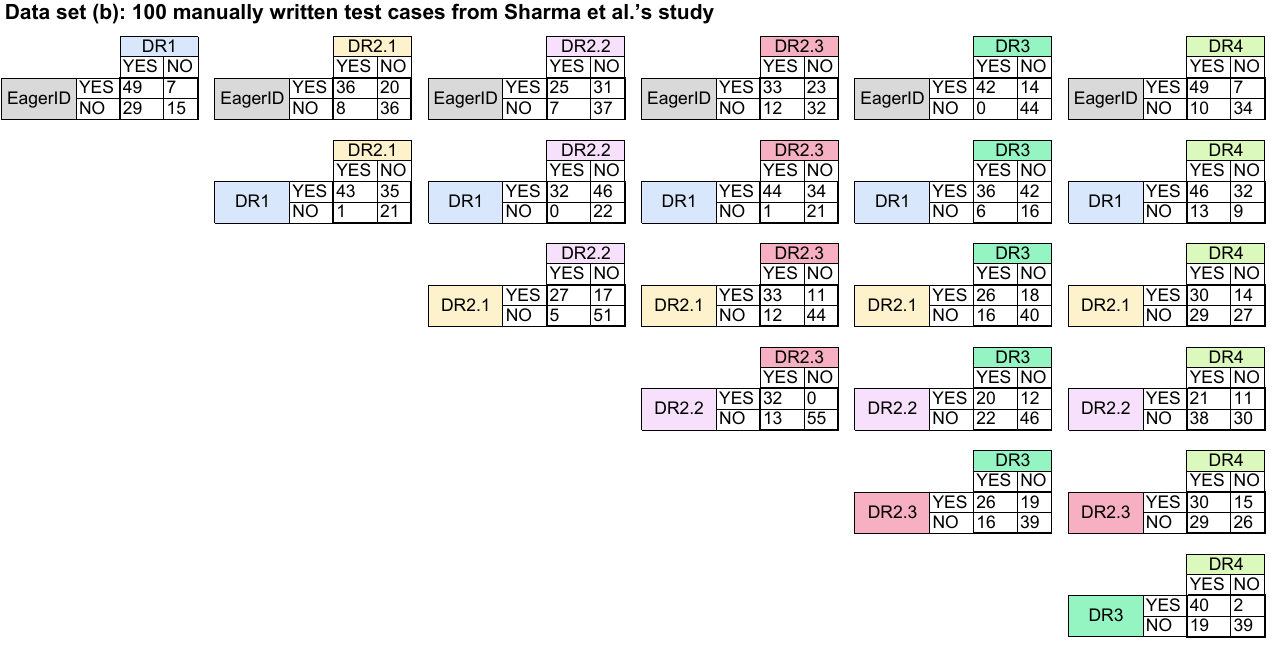}
\label{fig:ET_outcome_dataSetB}
\end{center}
\end{figure*}

\begin{figure*}[htb]
\begin{center}
\includegraphics[width=1\linewidth]{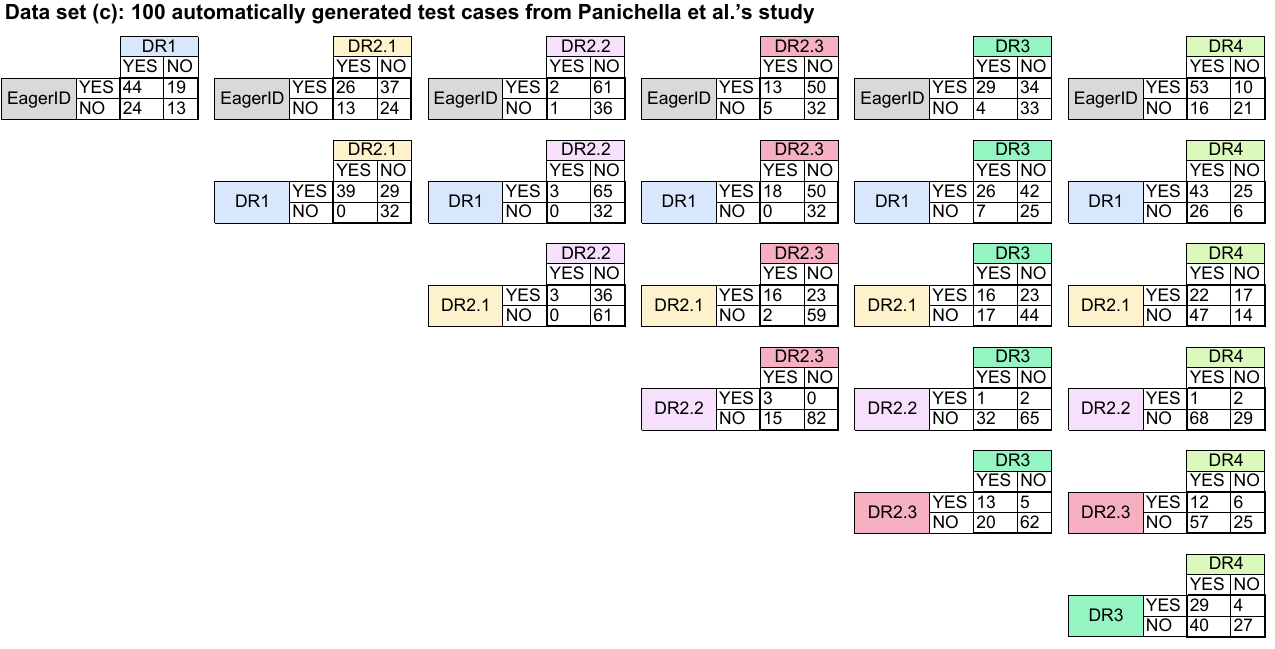}
\label{fig:ET_outcome_dataSetC}
\end{center}
\end{figure*}

\newpage\hbox{}\thispagestyle{empty}
\newpage
\newpage\hbox{}\thispagestyle{empty}
\newpage

\end{appendices}

\bibliographystyle{cas-model2-names}
\bibliography{jbib} 

\begin{thebibliography}{63}
\expandafter\ifx\csname natexlab\endcsname\relax\def\natexlab#1{#1}\fi
\providecommand{\url}[1]{\texttt{#1}}
\providecommand{\href}[2]{#2}
\providecommand{\path}[1]{#1}
\providecommand{\DOIprefix}{doi:}
\providecommand{\ArXivprefix}{arXiv:}
\providecommand{\URLprefix}{URL: }
\providecommand{\Pubmedprefix}{pmid:}
\providecommand{\doi}[1]{\href{http://dx.doi.org/#1}{\path{#1}}}
\providecommand{\Pubmed}[1]{\href{pmid:#1}{\path{#1}}}
\providecommand{\bibinfo}[2]{#2}
\ifx\xfnm\relax \def\xfnm[#1]{\unskip,\space#1}\fi
\bibitem[{Aljedaani et~al.(2021)Aljedaani, Peruma, Aljohani, Alotaibi, Mkaouer,
  Ouni, Newman, Ghallab and Ludi}]{Aljedaani2021Test}
\bibinfo{author}{Aljedaani, W.}, \bibinfo{author}{Peruma, A.},
  \bibinfo{author}{Aljohani, A.}, \bibinfo{author}{Alotaibi, M.},
  \bibinfo{author}{Mkaouer, M.W.}, \bibinfo{author}{Ouni, A.},
  \bibinfo{author}{Newman, C.D.}, \bibinfo{author}{Ghallab, A.},
  \bibinfo{author}{Ludi, S.}, \bibinfo{year}{2021}.
\newblock \bibinfo{title}{Test smell detection tools: A systematic mapping
  study}, in: \bibinfo{booktitle}{ACM International Conference Proceeding
  Series}, pp. \bibinfo{pages}{170--180}.
\bibitem[{Athanasiou et~al.(2014)Athanasiou, Nugroho, Visser and
  Zaidman}]{athanasiou2014test}
\bibinfo{author}{Athanasiou, D.}, \bibinfo{author}{Nugroho, A.},
  \bibinfo{author}{Visser, J.}, \bibinfo{author}{Zaidman, A.},
  \bibinfo{year}{2014}.
\newblock \bibinfo{title}{Test code quality and its relation to issue handling
  performance}.
\newblock \bibinfo{journal}{IEEE Transactions on Software Engineering}
  \bibinfo{volume}{40}, \bibinfo{pages}{1100--1125}.
\bibitem[{Bavota et~al.(2012)Bavota, Qusef, Oliveto, De~Lucia and
  Binkley}]{Bavota2012empirical}
\bibinfo{author}{Bavota, G.}, \bibinfo{author}{Qusef, A.},
  \bibinfo{author}{Oliveto, R.}, \bibinfo{author}{De~Lucia, A.},
  \bibinfo{author}{Binkley, D.}, \bibinfo{year}{2012}.
\newblock \bibinfo{title}{An empirical analysis of the distribution of unit
  test smells and their impact on software maintenance}, in:
  \bibinfo{booktitle}{IEEE International Conference on Software Maintenance,
  ICSM}, pp. \bibinfo{pages}{56--65}.
\bibitem[{Bavota et~al.(2015)Bavota, Qusef, Oliveto, De~Lucia and
  Binkley}]{Bavota2015Are}
\bibinfo{author}{Bavota, G.}, \bibinfo{author}{Qusef, A.},
  \bibinfo{author}{Oliveto, R.}, \bibinfo{author}{De~Lucia, A.},
  \bibinfo{author}{Binkley, D.}, \bibinfo{year}{2015}.
\newblock \bibinfo{title}{Are test smells really harmful? an empirical study}.
\newblock \bibinfo{journal}{Empirical Software Engineering}
  \bibinfo{volume}{20}, \bibinfo{pages}{1052--1094}.
\bibitem[{Beck(2003;2002;)}]{Beck2003Test}
\bibinfo{author}{Beck, K.}, \bibinfo{year}{2003;2002;}.
\newblock \bibinfo{title}{Test-driven development: by example}.
\newblock \bibinfo{publisher}{Addison-Wesley}, \bibinfo{address}{Boston, MA}.
\bibitem[{Binder(2000)}]{binder2000testing}
\bibinfo{author}{Binder, R.}, \bibinfo{year}{2000}.
\newblock \bibinfo{title}{Testing object-oriented systems: models, patterns,
  and tools}.
\newblock \bibinfo{publisher}{Addison-Wesley Professional}.
\bibitem[{Bowes et~al.(2017)Bowes, Hall, Petri{\'c}, Shippey and
  Turhan}]{bowes2017how}
\bibinfo{author}{Bowes, D.}, \bibinfo{author}{Hall, T.},
  \bibinfo{author}{Petri{\'c}, J.}, \bibinfo{author}{Shippey, T.},
  \bibinfo{author}{Turhan, B.}, \bibinfo{year}{2017}.
\newblock \bibinfo{title}{How good are my tests?}, in:
  \bibinfo{booktitle}{Proceedings of the International Workshop on Emerging
  Trends in Software Metrics, WETSoM}, pp. \bibinfo{pages}{9--14}.
\bibitem[{Breugelmans and Van~Rompaey(2008)}]{breugelmans2008testq}
\bibinfo{author}{Breugelmans, M.}, \bibinfo{author}{Van~Rompaey, B.},
  \bibinfo{year}{2008}.
\newblock \bibinfo{title}{Testq: Exploring structural and maintenance
  characteristics of unit test suites}, in: \bibinfo{booktitle}{WASDeTT-1: 1st
  International Workshop on Advanced Software Development Tools and
  Techniques}, \bibinfo{organization}{Citeseer}. p.~\bibinfo{pages}{11}.
\bibitem[{Camara et~al.(2021)Camara, Silva, Endo and Vergilio}]{Camara2021On}
\bibinfo{author}{Camara, B.}, \bibinfo{author}{Silva, M.},
  \bibinfo{author}{Endo, A.}, \bibinfo{author}{Vergilio, S.},
  \bibinfo{year}{2021}.
\newblock \bibinfo{title}{On the use of test smells for prediction of flaky
  tests}, in: \bibinfo{booktitle}{ACM International Conference Proceeding
  Series}, pp. \bibinfo{pages}{46--54}.
\bibitem[{Cohen(1960)}]{cohen1960coefficient}
\bibinfo{author}{Cohen, J.}, \bibinfo{year}{1960}.
\newblock \bibinfo{title}{A coefficient of agreement for nominal scales}.
\newblock \bibinfo{journal}{Educational and psychological measurement}
  \bibinfo{volume}{20}, \bibinfo{pages}{37--46}.
\bibitem[{Damasceno et~al.(2023)Damasceno, Bezerra, Coutinho and
  Machado}]{Damasceno2022Analyzing}
\bibinfo{author}{Damasceno, H.}, \bibinfo{author}{Bezerra, C.},
  \bibinfo{author}{Coutinho, E.}, \bibinfo{author}{Machado, I.},
  \bibinfo{year}{2023}.
\newblock \bibinfo{title}{Analyzing test smells refactoring from a developers
  perspective}, in: \bibinfo{booktitle}{Proceedings of the XXI Brazilian
  Symposium on Software Quality}, \bibinfo{publisher}{Association for Computing
  Machinery}, \bibinfo{address}{New York, NY, USA}.
\bibitem[{De~Bleser et~al.(2019a)De~Bleser, Di~Nucci and
  De~Roover}]{Bleser2019Assessing}
\bibinfo{author}{De~Bleser, J.}, \bibinfo{author}{Di~Nucci, D.},
  \bibinfo{author}{De~Roover, C.}, \bibinfo{year}{2019}a.
\newblock \bibinfo{title}{Assessing diffusion and perception of test smells in
  scala projects}, in: \bibinfo{booktitle}{IEEE International Working
  Conference on Mining Software Repositories}, pp. \bibinfo{pages}{457--467}.
\bibitem[{De~Bleser et~al.(2019b)De~Bleser, Di~Nucci and
  De~Roover}]{Bleser2019SoCRATES}
\bibinfo{author}{De~Bleser, J.}, \bibinfo{author}{Di~Nucci, D.},
  \bibinfo{author}{De~Roover, C.}, \bibinfo{year}{2019}b.
\newblock \bibinfo{title}{Socrates,: Scala radar for test smells}, in:
  \bibinfo{booktitle}{Scala 2019 - Proceedings of the 10th ACM SIGPLAN
  International Symposium on Scala, Part of ECOOP 2019}, pp.
  \bibinfo{pages}{22--26}.
\bibitem[{Dragan et~al.(2006)Dragan, Collard and Maletic}]{dragan2006reverse}
\bibinfo{author}{Dragan, N.}, \bibinfo{author}{Collard, M.L.},
  \bibinfo{author}{Maletic, J.I.}, \bibinfo{year}{2006}.
\newblock \bibinfo{title}{Reverse engineering method stereotypes}, in:
  \bibinfo{booktitle}{2006 22nd IEEE International Conference on Software
  Maintenance}, \bibinfo{organization}{IEEE}. pp. \bibinfo{pages}{24--34}.
\bibitem[{Dragan et~al.(2009)Dragan, Collard and Maletic}]{dragan2009using}
\bibinfo{author}{Dragan, N.}, \bibinfo{author}{Collard, M.L.},
  \bibinfo{author}{Maletic, J.I.}, \bibinfo{year}{2009}.
\newblock \bibinfo{title}{Using method stereotype distribution as a signature
  descriptor for software systems}, in: \bibinfo{booktitle}{2009 IEEE
  International Conference on Software Maintenance},
  \bibinfo{organization}{IEEE}. pp. \bibinfo{pages}{567--570}.
\bibitem[{Fowler(2018)}]{fowler2018refactoring}
\bibinfo{author}{Fowler, M.}, \bibinfo{year}{2018}.
\newblock \bibinfo{title}{Refactoring: improving the design of existing code}.
\newblock \bibinfo{publisher}{Addison-Wesley Professional}.
\bibitem[{Fraser and Arcuri(2014)}]{fraser2014large}
\bibinfo{author}{Fraser, G.}, \bibinfo{author}{Arcuri, A.},
  \bibinfo{year}{2014}.
\newblock \bibinfo{title}{A large-scale evaluation of automated unit test
  generation using evosuite}.
\newblock \bibinfo{journal}{ACM Transactions on Software Engineering and
  Methodology (TOSEM)} \bibinfo{volume}{24}, \bibinfo{pages}{1--42}.
\bibitem[{Freeman and Pryce(2009)}]{freeman2009growing}
\bibinfo{author}{Freeman, S.}, \bibinfo{author}{Pryce, N.},
  \bibinfo{year}{2009}.
\newblock \bibinfo{title}{Growing object-oriented software, guided by tests}.
\newblock \bibinfo{publisher}{Pearson Education}.
\bibitem[{Garousi and K{\"u}{\c{c}}{\"u}k(2018)}]{garousi2018smells}
\bibinfo{author}{Garousi, V.}, \bibinfo{author}{K{\"u}{\c{c}}{\"u}k, B.},
  \bibinfo{year}{2018}.
\newblock \bibinfo{title}{Smells in software test code: A survey of knowledge
  in industry and academia}.
\newblock \bibinfo{journal}{Journal of systems and software}
  \bibinfo{volume}{138}, \bibinfo{pages}{52--81}.
\bibitem[{Goodenough and Gerhart(1975)}]{goodenough1975toward}
\bibinfo{author}{Goodenough, J.B.}, \bibinfo{author}{Gerhart, S.L.},
  \bibinfo{year}{1975}.
\newblock \bibinfo{title}{Toward a theory of test data selection}.
\newblock \bibinfo{journal}{IEEE transactions on software engineering}
  \bibinfo{volume}{SE-1}, \bibinfo{pages}{156--173}.
\bibitem[{Grano et~al.(2020)Grano, De~Iaco, Palomba and Gall}]{grano2020pinsa}
\bibinfo{author}{Grano, G.}, \bibinfo{author}{De~Iaco, C.},
  \bibinfo{author}{Palomba, F.}, \bibinfo{author}{Gall, H.C.},
  \bibinfo{year}{2020}.
\newblock \bibinfo{title}{Pizza versus pinsa: On the perception and
  measurability of unit test code quality}, in: \bibinfo{booktitle}{Proceedings
  of the 36th IEEE International Conference on Software Maintenance and
  Evolution}, \bibinfo{publisher}{{IEEE}}. pp. \bibinfo{pages}{336--347}.
\bibitem[{Hamill(2004)}]{hamill2004unit}
\bibinfo{author}{Hamill, P.}, \bibinfo{year}{2004}.
\newblock \bibinfo{title}{Unit test frameworks: tools for high-quality software
  development}.
\newblock \bibinfo{publisher}{" O'Reilly Media, Inc."}.
\bibitem[{Jabbari et~al.(2018)Jabbari, Ali, Petersen and
  Tanveer}]{JabbariAPT18}
\bibinfo{author}{Jabbari, R.}, \bibinfo{author}{Ali, N.B.},
  \bibinfo{author}{Petersen, K.}, \bibinfo{author}{Tanveer, B.},
  \bibinfo{year}{2018}.
\newblock \bibinfo{title}{Towards a benefits dependency network for devops
  based on a systematic literature review}.
\newblock \bibinfo{journal}{J. Softw. Evol. Process.} \bibinfo{volume}{30}.
\bibitem[{Junior et~al.(2020)Junior, Rocha, Martins and
  Machado}]{junior2020survey}
\bibinfo{author}{Junior, N.S.}, \bibinfo{author}{Rocha, L.},
  \bibinfo{author}{Martins, L.A.}, \bibinfo{author}{Machado, I.},
  \bibinfo{year}{2020}.
\newblock \bibinfo{title}{A survey on test practitioners' awareness of test
  smells}.
\newblock \bibinfo{journal}{arXiv preprint arXiv:2003.05613} .
\bibitem[{Kapfhammer and Soffa(2003)}]{kapfhammer2003family}
\bibinfo{author}{Kapfhammer, G.M.}, \bibinfo{author}{Soffa, M.L.},
  \bibinfo{year}{2003}.
\newblock \bibinfo{title}{A family of test adequacy criteria for
  database-driven applications}.
\newblock \bibinfo{journal}{ACM SIGSOFT Software Engineering Notes}
  \bibinfo{volume}{28}, \bibinfo{pages}{98--107}.
\bibitem[{Khorikov(2020)}]{khorikov2020unit}
\bibinfo{author}{Khorikov, V.}, \bibinfo{year}{2020}.
\newblock \bibinfo{title}{Unit Testing Principles, Practices, and Patterns}.
\newblock \bibinfo{publisher}{Simon and Schuster}.
\bibitem[{Kochhar et~al.(2019)Kochhar, Xia and Lo}]{Kochhar2019Practitioners}
\bibinfo{author}{Kochhar, P.S.}, \bibinfo{author}{Xia, X.},
  \bibinfo{author}{Lo, D.}, \bibinfo{year}{2019}.
\newblock \bibinfo{title}{Practitioners' views on good software testing
  practices}, in: \bibinfo{booktitle}{Proceedings - 2019 IEEE/ACM 41st
  International Conference on Software Engineering: Software Engineering in
  Practice, ICSE-SEIP 2019}, pp. \bibinfo{pages}{61--70}.
\bibitem[{Lambiase et~al.(2020)Lambiase, Cupito, Pecorelli, De~Lucia and
  Palomba}]{Lambiase2020Just}
\bibinfo{author}{Lambiase, S.}, \bibinfo{author}{Cupito, A.},
  \bibinfo{author}{Pecorelli, F.}, \bibinfo{author}{De~Lucia, A.},
  \bibinfo{author}{Palomba, F.}, \bibinfo{year}{2020}.
\newblock \bibinfo{title}{Just-in-time test smell detection and refactoring:
  The darts project}, in: \bibinfo{booktitle}{IEEE International Conference on
  Program Comprehension}, pp. \bibinfo{pages}{441--445}.
\bibitem[{Landis and Koch(1977)}]{landis1977measurement}
\bibinfo{author}{Landis, J.R.}, \bibinfo{author}{Koch, G.G.},
  \bibinfo{year}{1977}.
\newblock \bibinfo{title}{The measurement of observer agreement for categorical
  data}.
\newblock \bibinfo{journal}{Biometrics} \bibinfo{volume}{33},
  \bibinfo{pages}{159--174}.
\bibitem[{Lemos et~al.(2007)Lemos, Vincenzi, Maldonado and
  Masiero}]{lemos2007control}
\bibinfo{author}{Lemos, O.A.L.}, \bibinfo{author}{Vincenzi, A.M.R.},
  \bibinfo{author}{Maldonado, J.C.}, \bibinfo{author}{Masiero, P.C.},
  \bibinfo{year}{2007}.
\newblock \bibinfo{title}{Control and data flow structural testing criteria for
  aspect-oriented programs}.
\newblock \bibinfo{journal}{Journal of Systems and Software}
  \bibinfo{volume}{80}, \bibinfo{pages}{862--882}.
\bibitem[{Lilis and Savidis(2019)}]{lilis2019survey}
\bibinfo{author}{Lilis, Y.}, \bibinfo{author}{Savidis, A.},
  \bibinfo{year}{2019}.
\newblock \bibinfo{title}{A survey of metaprogramming languages}.
\newblock \bibinfo{journal}{ACM Computing Surveys (CSUR)} \bibinfo{volume}{52},
  \bibinfo{pages}{1--39}.
\bibitem[{Martin(2014)}]{fowler2014unit}
\bibinfo{author}{Martin, F.}, \bibinfo{year}{2014}.
\newblock \bibinfo{title}{Unit test}.
\newblock \URLprefix \url{https://martinfowler.com/bliki/UnitTest.html}.
\bibitem[{Martins et~al.(2023)Martins, Costa and Machado}]{Martins2023On}
\bibinfo{author}{Martins, L.}, \bibinfo{author}{Costa, H.},
  \bibinfo{author}{Machado, I.}, \bibinfo{year}{2023}.
\newblock \bibinfo{title}{On the diffusion of test smells and their
  relationship with test code quality of java projects}.
\newblock \bibinfo{journal}{Journal of Software: Evolution and Process} .
\bibitem[{Meszaros(2007)}]{meszaros2007xunit}
\bibinfo{author}{Meszaros, G.}, \bibinfo{year}{2007}.
\newblock \bibinfo{title}{xUnit Test Patterns: Refactoring Test Code}.
\newblock \bibinfo{publisher}{Addison-Wesley Professional}.
\bibitem[{Moreno and Marcus(2012)}]{moreno2012jstereocode}
\bibinfo{author}{Moreno, L.}, \bibinfo{author}{Marcus, A.},
  \bibinfo{year}{2012}.
\newblock \bibinfo{title}{Jstereocode: automatically identifying method and
  class stereotypes in java code}, in: \bibinfo{booktitle}{Proceedings of the
  27th IEEE/ACM International Conference on Automated Software Engineering},
  pp. \bibinfo{pages}{358--361}.
\bibitem[{Munaiah et~al.(2017)Munaiah, Kroh, Cabrey and
  Nagappan}]{munaiah2017curating}
\bibinfo{author}{Munaiah, N.}, \bibinfo{author}{Kroh, S.},
  \bibinfo{author}{Cabrey, C.}, \bibinfo{author}{Nagappan, M.},
  \bibinfo{year}{2017}.
\newblock \bibinfo{title}{Curating github for engineered software projects}.
\newblock \bibinfo{journal}{Empirical Software Engineering}
  \bibinfo{volume}{22}, \bibinfo{pages}{3219--3253}.
\bibitem[{Neukirchen et~al.(2008)Neukirchen, Zeiss and
  Grabowski}]{neukirchen2008approach}
\bibinfo{author}{Neukirchen, H.}, \bibinfo{author}{Zeiss, B.},
  \bibinfo{author}{Grabowski, J.}, \bibinfo{year}{2008}.
\newblock \bibinfo{title}{An approach to quality engineering of {TTCN}-3 test
  specifications}.
\newblock \bibinfo{journal}{International Journal on Software Tools for
  Technology Transfer} \bibinfo{volume}{10}, \bibinfo{pages}{309}.
\bibitem[{Palomba et~al.(2018)Palomba, Zaidman and
  De~Lucia}]{Palomba2018Automatic}
\bibinfo{author}{Palomba, F.}, \bibinfo{author}{Zaidman, A.},
  \bibinfo{author}{De~Lucia, A.}, \bibinfo{year}{2018}.
\newblock \bibinfo{title}{Automatic test smell detection using information
  retrieval techniques}, in: \bibinfo{booktitle}{Proceedings - 2018 IEEE
  International Conference on Software Maintenance and Evolution, ICSME 2018},
  pp. \bibinfo{pages}{311--322}.
\bibitem[{Panichella et~al.(2020)Panichella, Panichella, Fraser, Sawant and
  Hellendoorn}]{Panichella2020Revisiting}
\bibinfo{author}{Panichella, A.}, \bibinfo{author}{Panichella, S.},
  \bibinfo{author}{Fraser, G.}, \bibinfo{author}{Sawant, A.A.},
  \bibinfo{author}{Hellendoorn, V.J.}, \bibinfo{year}{2020}.
\newblock \bibinfo{title}{Revisiting test smells in automatically generated
  tests: Limitations, pitfalls, and opportunities}, in:
  \bibinfo{booktitle}{2020 IEEE International Conference on Software
  Maintenance and Evolution (ICSME)}, pp. \bibinfo{pages}{523--533}.
\bibitem[{Panichella et~al.(2022)Panichella, Panichella, Fraser, Sawant and
  Hellendoorn}]{panichella2022test}
\bibinfo{author}{Panichella, A.}, \bibinfo{author}{Panichella, S.},
  \bibinfo{author}{Fraser, G.}, \bibinfo{author}{Sawant, A.A.},
  \bibinfo{author}{Hellendoorn, V.J.}, \bibinfo{year}{2022}.
\newblock \bibinfo{title}{Test smells 20 years later: detectability, validity,
  and reliability}.
\newblock \bibinfo{journal}{Empirical Software Engineering}
  \bibinfo{volume}{27}.
\bibitem[{Pei et~al.(2017)Pei, Cao, Yang and Jana}]{pei2019deepxplore}
\bibinfo{author}{Pei, K.}, \bibinfo{author}{Cao, Y.}, \bibinfo{author}{Yang,
  J.}, \bibinfo{author}{Jana, S.}, \bibinfo{year}{2017}.
\newblock \bibinfo{title}{Deepxplore: Automated whitebox testing of deep
  learning systems}, in: \bibinfo{booktitle}{Proceedings of the 26th ACM
  Symposium on Operating Systems Principles - {SOSP}}, pp.
  \bibinfo{pages}{1--18}.
\bibitem[{Peruma et~al.(2020)Peruma, Almalki, Newman, Mkaouer, Ouni and
  Palomba}]{Peruma2020TsDetect}
\bibinfo{author}{Peruma, A.}, \bibinfo{author}{Almalki, K.},
  \bibinfo{author}{Newman, C.D.}, \bibinfo{author}{Mkaouer, M.W.},
  \bibinfo{author}{Ouni, A.}, \bibinfo{author}{Palomba, F.},
  \bibinfo{year}{2020}.
\newblock \bibinfo{title}{Tsdetect: An open source test smells detection tool},
  in: \bibinfo{booktitle}{ESEC/FSE 2020 - Proceedings of the 28th ACM Joint
  Meeting European Software Engineering Conference and Symposium on the
  Foundations of Software Engineering}, pp. \bibinfo{pages}{1650--1654}.
\bibitem[{Pizzini et~al.(2023)Pizzini, Reinehr and
  Malucelli}]{Pizzini2022Automatic}
\bibinfo{author}{Pizzini, A.}, \bibinfo{author}{Reinehr, S.},
  \bibinfo{author}{Malucelli, A.}, \bibinfo{year}{2023}.
\newblock \bibinfo{title}{Automatic refactoring method to remove eager test
  smell}, in: \bibinfo{booktitle}{Proceedings of the XXI Brazilian Symposium on
  Software Quality}, \bibinfo{publisher}{Association for Computing Machinery},
  \bibinfo{address}{New York, NY, USA}.
\bibitem[{Rwemalika et~al.(2023)Rwemalika, Habchi, Papadakis, Le~Traon and
  Brasseur}]{rwemalika2023smells}
\bibinfo{author}{Rwemalika, R.}, \bibinfo{author}{Habchi, S.},
  \bibinfo{author}{Papadakis, M.}, \bibinfo{author}{Le~Traon, Y.},
  \bibinfo{author}{Brasseur, M..}, \bibinfo{year}{2023}.
\newblock \bibinfo{title}{Smells in system user interactive tests}.
\newblock \bibinfo{journal}{Empirical Software Engineering}
  \bibinfo{volume}{28}.
\bibitem[{Santana et~al.(2022)Santana, Martins, Virgínio, Soares, Costa and
  Machado}]{Santana2022Refactoring}
\bibinfo{author}{Santana, R.}, \bibinfo{author}{Martins, L.},
  \bibinfo{author}{Virgínio, T.}, \bibinfo{author}{Soares, L.},
  \bibinfo{author}{Costa, H.}, \bibinfo{author}{Machado, I.},
  \bibinfo{year}{2022}.
\newblock \bibinfo{title}{Refactoring assertion roulette and duplicate assert
  test smells: a controlled experiment}, in: \bibinfo{booktitle}{CIbSE 2022 -
  XXV Ibero-American Conference on Software Engineering}.
\bibitem[{Schvarcbacher et~al.(2019)Schvarcbacher, Spadini, Bruntink and
  Oprescu}]{Schvarcbacher2019Investigating}
\bibinfo{author}{Schvarcbacher, M.}, \bibinfo{author}{Spadini, D.},
  \bibinfo{author}{Bruntink, M.}, \bibinfo{author}{Oprescu, A.},
  \bibinfo{year}{2019}.
\newblock \bibinfo{title}{Investigating developer perception on test smells
  using better code hub - work in progress -}, in: \bibinfo{booktitle}{CEUR
  Workshop Proceedings}.
\bibitem[{Sharma et~al.(2023)Sharma, Georgiou, Kechagia, Ghaleb and
  Sarro}]{tushar2023investigating}
\bibinfo{author}{Sharma, T.}, \bibinfo{author}{Georgiou, S.},
  \bibinfo{author}{Kechagia, M.}, \bibinfo{author}{Ghaleb, T.A.},
  \bibinfo{author}{Sarro, F.}, \bibinfo{year}{2023}.
\newblock \bibinfo{title}{Investigating developers’ perception on software
  testability and its effects}.
\newblock \bibinfo{journal}{Empirical software engineering : an international
  journal} \bibinfo{volume}{28}, \bibinfo{pages}{120}.
\bibitem[{Spadini et~al.(2020)Spadini, Schvarcbacher, Oprescu, Bruntink and
  Bacchelli}]{Spadini2020Investigating}
\bibinfo{author}{Spadini, D.}, \bibinfo{author}{Schvarcbacher, M.},
  \bibinfo{author}{Oprescu, A..}, \bibinfo{author}{Bruntink, M.},
  \bibinfo{author}{Bacchelli, A.}, \bibinfo{year}{2020}.
\newblock \bibinfo{title}{Investigating severity thresholds for test smells},
  in: \bibinfo{booktitle}{Proceedings - 2020 IEEE/ACM 17th International
  Conference on Mining Software Repositories, MSR 2020}, pp.
  \bibinfo{pages}{311--321}.
\bibitem[{Spadini2018On et~al.(2018)Spadini2018On, Palomba, Zaidman, Bruntink
  and Bacchelli}]{Spadini2018On}
\bibinfo{author}{Spadini2018On, D.}, \bibinfo{author}{Palomba, F.},
  \bibinfo{author}{Zaidman, A.}, \bibinfo{author}{Bruntink, M.},
  \bibinfo{author}{Bacchelli, A.}, \bibinfo{year}{2018}.
\newblock \bibinfo{title}{On the relation of test smells to software code
  quality}, in: \bibinfo{booktitle}{Proceedings - 2018 IEEE International
  Conference on Software Maintenance and Evolution, ICSME 2018}, pp.
  \bibinfo{pages}{1--12}.
\bibitem[{Tahir et~al.(2016)Tahir, Counsell and MacDonell}]{Tahir2016Empirical}
\bibinfo{author}{Tahir, A.}, \bibinfo{author}{Counsell, S.},
  \bibinfo{author}{MacDonell, S.G.}, \bibinfo{year}{2016}.
\newblock \bibinfo{title}{An empirical study into the relationship between
  class features and test smells}, in: \bibinfo{booktitle}{Proceedings -
  Asia-Pacific Software Engineering Conference, APSEC}, pp.
  \bibinfo{pages}{137--144}.
\bibitem[{Tran et~al.(2019)Tran, Ali, B{\"o}rstler and
  Unterkalmsteiner}]{tran2019test}
\bibinfo{author}{Tran, H.K.V.}, \bibinfo{author}{Ali, N.B.},
  \bibinfo{author}{B{\"o}rstler, J.}, \bibinfo{author}{Unterkalmsteiner, M.},
  \bibinfo{year}{2019}.
\newblock \bibinfo{title}{Test-{Case} {Quality} -- {Understanding}
  {Practitioners}' {Perspectives}}, in: \bibinfo{booktitle}{Proceedings of the
  20th {International} {Conference} on {Product}-{Focused} {Software} {Process}
  {Improvement} ({PROFES})}, \bibinfo{publisher}{Springer}. pp.
  \bibinfo{pages}{37--52}.
\bibitem[{Tran et~al.(2021)Tran, Unterkalmsteiner, B{\"o}rstler and
  Ali}]{tran2021assessing}
\bibinfo{author}{Tran, H.K.V.}, \bibinfo{author}{Unterkalmsteiner, M.},
  \bibinfo{author}{B{\"o}rstler, J.}, \bibinfo{author}{Ali, N.B.},
  \bibinfo{year}{2021}.
\newblock \bibinfo{title}{Assessing test artifact quality---a tertiary study}.
\newblock \bibinfo{journal}{Information and Software Technology}
  \bibinfo{volume}{139}, \bibinfo{pages}{106620}.
\bibitem[{Tudose(2021;2020;)}]{Tudose2021JUnit}
\bibinfo{author}{Tudose, C.}, \bibinfo{year}{2021;2020;}.
\newblock \bibinfo{title}{JUnit in Action, Third Edition}.
\newblock \bibinfo{edition}{Third} ed., \bibinfo{publisher}{Manning
  Publications}.
\bibitem[{Tufano et~al.(2016a)Tufano, Palomba, Bavota, Di~Penta, Oliveto,
  De~Lucia and Poshyvanyk}]{Tufano2016An}
\bibinfo{author}{Tufano, M.}, \bibinfo{author}{Palomba, F.},
  \bibinfo{author}{Bavota, G.}, \bibinfo{author}{Di~Penta, M.},
  \bibinfo{author}{Oliveto, R.}, \bibinfo{author}{De~Lucia, A.},
  \bibinfo{author}{Poshyvanyk, D.}, \bibinfo{year}{2016}a.
\newblock \bibinfo{title}{An empirical investigation into the nature of test
  smells}, in: \bibinfo{booktitle}{ASE 2016 - Proceedings of the 31st IEEE/ACM
  International Conference on Automated Software Engineering}, pp.
  \bibinfo{pages}{4--15}.
\bibitem[{Tufano et~al.(2016b)Tufano, Palomba, Bavota, Di~Penta, Oliveto,
  De~Lucia and Poshyvanyk}]{Tufano2016empirical}
\bibinfo{author}{Tufano, M.}, \bibinfo{author}{Palomba, F.},
  \bibinfo{author}{Bavota, G.}, \bibinfo{author}{Di~Penta, M.},
  \bibinfo{author}{Oliveto, R.}, \bibinfo{author}{De~Lucia, A.},
  \bibinfo{author}{Poshyvanyk, D.}, \bibinfo{year}{2016}b.
\newblock \bibinfo{title}{An empirical investigation into the nature of test
  smells}, in: \bibinfo{booktitle}{ASE 2016 - Proceedings of the 31st IEEE/ACM
  International Conference on Automated Software Engineering}, pp.
  \bibinfo{pages}{4--15}.
\bibitem[{Tufano et~al.(2015)Tufano, Palomba, Bavota, Oliveto, Di~Penta,
  De~Lucia and Poshyvanyk}]{tufano2015and}
\bibinfo{author}{Tufano, M.}, \bibinfo{author}{Palomba, F.},
  \bibinfo{author}{Bavota, G.}, \bibinfo{author}{Oliveto, R.},
  \bibinfo{author}{Di~Penta, M.}, \bibinfo{author}{De~Lucia, A.},
  \bibinfo{author}{Poshyvanyk, D.}, \bibinfo{year}{2015}.
\newblock \bibinfo{title}{When and why your code starts to smell bad}, in:
  \bibinfo{booktitle}{2015 IEEE/ACM 37th IEEE International Conference on
  Software Engineering}, \bibinfo{organization}{IEEE}. pp.
  \bibinfo{pages}{403--414}.
\bibitem[{Van~Deursen et~al.(2001)Van~Deursen, Moonen, Van Den~Bergh and
  Kok}]{van2001refactoring}
\bibinfo{author}{Van~Deursen, A.}, \bibinfo{author}{Moonen, L.},
  \bibinfo{author}{Van Den~Bergh, A.}, \bibinfo{author}{Kok, G.},
  \bibinfo{year}{2001}.
\newblock \bibinfo{title}{Refactoring test code}, in:
  \bibinfo{booktitle}{Proceedings of the 2nd international conference on
  extreme programming and flexible processes in software engineering (XP2001)},
  \bibinfo{organization}{Citeseer}. pp. \bibinfo{pages}{92--95}.
\bibitem[{Van~Rompaey et~al.(2006)Van~Rompaey, Du~Bois and
  Demeyer}]{Van2006Characterizing}
\bibinfo{author}{Van~Rompaey, B.}, \bibinfo{author}{Du~Bois, B.},
  \bibinfo{author}{Demeyer, S.}, \bibinfo{year}{2006}.
\newblock \bibinfo{title}{Characterizing the relative significance of a test
  smell}, in: \bibinfo{booktitle}{IEEE International Conference on Software
  Maintenance, ICSM}, pp. \bibinfo{pages}{391--400}.
\bibitem[{Van~Rompaey et~al.(2007)Van~Rompaey, Du~Bois, Demeyer and
  Rieger}]{Rompaey2007On}
\bibinfo{author}{Van~Rompaey, B.}, \bibinfo{author}{Du~Bois, B.},
  \bibinfo{author}{Demeyer, S.}, \bibinfo{author}{Rieger, M.},
  \bibinfo{year}{2007}.
\newblock \bibinfo{title}{On the detection of test smells: A metrics-based
  approach for general fixture and eager test}.
\newblock \bibinfo{journal}{IEEE Transactions on Software Engineering}
  \bibinfo{volume}{33}, \bibinfo{pages}{800--816}.
\bibitem[{Virg{\'\i}nio et~al.(2020)Virg{\'\i}nio, Martins, Rocha, Santana,
  Cruz, Costa and Machado}]{virginio2020jnose}
\bibinfo{author}{Virg{\'\i}nio, T.}, \bibinfo{author}{Martins, L.},
  \bibinfo{author}{Rocha, L.}, \bibinfo{author}{Santana, R.},
  \bibinfo{author}{Cruz, A.}, \bibinfo{author}{Costa, H.},
  \bibinfo{author}{Machado, I.}, \bibinfo{year}{2020}.
\newblock \bibinfo{title}{Jnose: Java test smell detector}, in:
  \bibinfo{booktitle}{Proceedings of the XXXIV Brazilian Symposium on Software
  Engineering}, pp. \bibinfo{pages}{564--569}.
\bibitem[{Virg{\'\i}nio et~al.(2019)Virg{\'\i}nio, Soares, Santana, Costa,
  Martins and Machado}]{Virginio2019On}
\bibinfo{author}{Virg{\'\i}nio, T.}, \bibinfo{author}{Soares, L.R.},
  \bibinfo{author}{Santana, R.}, \bibinfo{author}{Costa, H.},
  \bibinfo{author}{Martins, L.A.}, \bibinfo{author}{Machado, I.},
  \bibinfo{year}{2019}.
\newblock \bibinfo{title}{On the influence of test smells on test coverage},
  in: \bibinfo{booktitle}{ACM International Conference Proceeding Series}, pp.
  \bibinfo{pages}{467--471}.
\bibitem[{White and Krinke(2022)}]{white2022TCTracer}
\bibinfo{author}{White, R.}, \bibinfo{author}{Krinke, J.},
  \bibinfo{year}{2022}.
\newblock \bibinfo{title}{Tctracer: Establishing test-to-code traceability
  links using dynamic and static techniques}.
\newblock \bibinfo{journal}{Empirical Software Engineering}
  \bibinfo{volume}{27}.
\bibitem[{Zhu et~al.(1997)Zhu, Hall and May}]{zhu1997software}
\bibinfo{author}{Zhu, H.}, \bibinfo{author}{Hall, P.A.}, \bibinfo{author}{May,
  J.H.}, \bibinfo{year}{1997}.
\newblock \bibinfo{title}{Software unit test coverage and adequacy}.
\newblock \bibinfo{journal}{{ACM} computing surveys} \bibinfo{volume}{29},
  \bibinfo{pages}{366--427}.

\end{thebibliography}

\end{document}